\documentclass[aps,prb,twocolumn,showpacs,amsmath,floatfix]{revtex4}
\usepackage{graphicx}
\usepackage{dcolumn}
\usepackage{subfigure}
\usepackage{wrapfig}
\usepackage{cancel}
\usepackage{color}
\usepackage{bm}
\usepackage{ulem}
\usepackage{verbatim}
\usepackage{multirow}

\begin{document}

\def\k{{\bf k}}
\def\rr{{\bf r}}
\def\q{{\bf q}}
\def\v{{\bf v}}
\newcommand{\blue}{\textcolor{blue}}
\newcommand{\red}{\textcolor{red}}
\newcommand{\green}{\textcolor{green}}

\newcommand{\BaNiAs}{BaNi$_{\textnormal{2}}$As$_{\textnormal{2}} $}
\newcommand{\SrFeAs}{SrFe$_{\textnormal{2}}$As$_{\textnormal{2}} $}
\newcommand{\SrBaFeAs}{(Sr,Ba)Fe$_{\textnormal{2}}$As$_{\textnormal{2}}$}
\newcommand{\SrNiP}{SrNi$_{\textnormal{2}}$P$_{\textnormal{2}} $}
\newcommand{\SrKFeAs}{Sr$_{\textnormal{1-x}}$K$_{\textnormal{x}}$Fe$_{\textnormal{2}}$As$_{\textnormal{2}} $}
\newcommand{\pipi}{($\pi$,$\pi$)\ }
\newcommand{\BFCA}{Ba(Fe$_{1-x}$Co$_x$)$_2$As$_2$}
\newcommand{\BFA}{BaFe$_2$As$_2$}
\newcommand{\BFNA}{Ba(Fe$_{1-x}$Ni$_x$)$_2$As$_2$}
\newcommand{\BFKA}{BaFe$_2$(As$_{1-x}$K$_x$)$_2$}
\newcommand{\LFPO}{LaFePO}
\newcommand{\KFA}{KFe$_2$As$_2$}
\newcommand{\BFAP}{BaFe$_2$(As$_{1-x}$P$_x$)$_2$}
\newcommand{\etal}{{\it et al.}}

\title{\bf  Transport properties of 3D extended $s$-wave states in Fe-based superconductors}
\author{V. Mishra$^1$, S. Graser$^2$ and  P. J. Hirschfeld$^1$}
\affiliation{$^1$Department of Physics, University of Florida,
Gainesville, FL 32611, USA \\$^2$
Center for Electronic Correlations and Magnetism, Institute of Physics,
University of Augsburg, D-86135 Augsburg, Germany
}
\date{\today}

\begin{abstract}
The Fermi surfaces of Fe-pnictide superconductors are fairly two-dimensional (2D), and it has thus come as
a surprise that recent  penetration depth and thermal conductivity measurements on  systems of the 122 type
have reported $c$-axis  transport coefficients at low temperatures in the superconducting state comparable to or
even larger than that in  the $ab$-plane.  These results should provide important information on both the Fermi
surface and the superconducting state.  Here we consider the theory of the superfluid density and thermal
conductivity in models of extended-$s$ wave superconducting states expected to be appropriate for Fe-pnictide systems.
We include  intraband disorder and consider a range of different Fermi surfaces where gap nodes might exist.
We show that  recent experiments on  Ba(Fe$_{1-x}$Co$_x$)$_2$As$_2$ can be semiquantitatively understood by such an approach,
and discuss their implications.
\end{abstract}

\maketitle

\section{Introduction}

\label{sec:intro}

The new  Fe-based superconductors have captured the imagination of the theoretical and
experimental superconductivity community, in part because the critical temperature $T_c$ is
high in certain materials, but also  because the phenomenology of the superconducting
state appears to be quite different from any class of novel superconductors heretofore
discovered \cite{Ishida2009,Johnston2010}.  Since calculations of the coupling strength
$\lambda_{el-ph}$ due to the electron-phonon interaction show it to be small \cite{Boeri2008,Boeri2010,Stojchevska2010},
it is believed that  electronic pairing is predominant; thus attention has  focussed on sign-changing states with $A_{1g}$
symmetry predicted by spin fluctuation theory \cite{i_mazin_08,s_graser_08,a_chubukov_08,a_chubukov_09,k_kuroki_08,wang2009,wang2010}.
Depending on the details of the microscopic pair interaction\cite{a_chubukov_09,r_thomale_09,a_kemper_10} such states may display
``accidental" nodal structures, nodes not dictated by symmetry considerations.
It is important to understand such structures, as they provide clues to the origin of pairing.

 In the \LFPO, \KFA, and \BFAP~systems there is considerable evidence for low-energy excitations
which could be produced by order parameters with nodes, but the structure of the nodal manifold
is still controversial.  In the heavily-studied \BFCA, \BFNA~and \BFKA~systems, which we will refer
to collectively as Ba-122, measurements have been interpreted variously in terms of  fully gapped behavior,
deep minima in the superconducting gap, or gap nodes, and there is some evidence that these conclusions may
be quite doping-dependent.  The experimental probe which has thus far provided the lowest temperature information
on the bulk order parameter is thermal conductivity.  In the 122 systems, the  $ab$-plane thermal conductivity data
exhibited zero or extremely small linear-$T$ term in zero magnetic field, reflecting the apparent absence of any
nodes in the superconducting gap.  The field dependence, however, was significantly stronger than that expected
for a large-gap superconductor\cite{x_luo_09,Ding2009,Tanatar2010}.  This was analyzed by Mishra \etal\cite{v_mishra_09}
in terms of a gap with A$_{1g}$ symmetry with no nodes but deep minima on the electron sheets. {Bang proposed
that this effect} could also be explained phenomenologically by an isotropic $A_{1g}$ state with very small gap on one
Fermi surface sheet\cite{Bang2010}.

Recently, Martin \etal\cite{c_martin_10} reported a strong linear-$T$ behavior of the magnetic penetration depth
for currents along the $c$-axis in overdoped \BFNA crystals, compared to a predominantly $T^2$ behavior
in the $ab$-plane, with possible much smaller $ab$ linear-$T$ contributions for some dopings.  These authors
pointed to a need for theoretical analysis of superconductivity in these materials beyond 2D models.
Subsequently Reid \etal\cite{Reid2010} measured a  significant linear-$T$ term in the low-$T$ $c$-axis thermal
conductivity of \BFCA crystals, compared to a smaller or zero linear term in the $ab$-plane.  In addition, the
anisotropy ratio $(\kappa_c/\kappa_{c,N})/(\kappa_a/\kappa_{a,N})$ of the electronic thermal conductivities
normalized to their values at $T_c$ determined via resistivity measurements and the Wiedemann-Franz law, was
found to increase rapidly as the crystal was doped away from optimal $T_c$.  These authors argued  that
such an anisotropy ratio could arise only from gap nodes located precisely at flared regions
of the quasi-cylindrical Fermi surfaces where $v_{F,c}/v_{F,a}\gg 1$,
such that the corresponding $\kappa_a$ arising at low $T$ from the nodal structures would
be negligible.

For some samples,
Reid \etal~reported a nonzero limiting value of
$\kappa_{ab}(T)/T$ of order 1 $\mu $W/K$^2$cm.  It is important to note that these values are
much smaller than the value predicted in the simple BCS theory of quasiparticle transport in
a nodal  superconductor, of order ${\kappa/T} \approx {N_0 v_F^2/(k_F v_\Delta)}$, where $N_0$ is the Fermi
level density of states, $v_F$ is the Fermi velocity and $v_\Delta$ is the gap velocity
$v_\Delta\equiv \partial \Delta_\k/\partial \k|_\mathrm{node}$. This value is also much smaller than that  observed in
the cuprate case\cite{Taillefer1997}.   The low-$T$ linear term in the thermal conductivity is expected to be
universal (disorder-independent for weak disorder)  in the $p$-wave or $d$-wave case\cite{m_graf_96,p_hirschfeld_96}.
Mishra \etal\cite{v_mishra_09} showed recently that this expression continues to hold in the extended-$s$
type states thought to be characteristic of Fe-pnictide systems, but that universality breaks down effectively due to the
strong dependence of $v_\Delta$ on disorder. 

Theory has also made some recent progress in dealing with deviations from pure 2D behavior.
Early spin {fluctuation} calculations for the pairing state of the Fe-pnictide materials \cite{i_mazin_08,s_graser_08,k_kuroki_08,f_wang_09}
found that, depending on details of electronic structure and interaction parameters,
nodes could occur on the electron sheets, but the gap on the hole sheets was always fairly large and  isotropic. On the other hand,
Graser \etal\cite{s_graser_10} recently presented a calculation of spin-fluctuation pairing in the 122 systems based on an RPA
treatment of a 5-orbital model derived from density functional theory (DFT)
electronic structure.  In some cases, particularly for substantial hole doping, the ground state
was found to be of $A_{1g}$ symmetry, as in the 2D case, but while the electron sheet gap was found to
be highly anisotropic but nodeless, the hole sheet was found to have nodes near the top of the Brillouin
zone near the point where the Fermi surface sheets {experience} some outward flaring.  Similar results were
subsequently obtained by Kuroki \etal~for \BFAP \cite{Suzuki2010}.

In this work, we consider various possible Fermi surface geometries and gap structures which may give rise
to the unusual low transport anisotropy seen in Refs. \onlinecite{c_martin_09,Reid2010}, bearing in mind that any
phenomenology which purports to explain the results of these works must also be consistent with
the earlier thermal conductivity results of Refs. \onlinecite{x_luo_09,Ding2009,Tanatar2010}.
Although we consider 3D order parameter structures of the type found in Graser \etal\cite{s_graser_10},
we do not attempt to tie our calculations or parameters to any particular microscopic
calculation of either the Fermi surface or gap function, but rather to place restrictions on
what types of structures are possible on the basis of these measurements.   We argue that
gap structures with small nodal segments near the $k_z=\pm \pi$ sections of flared Fermi surfaces
are the most likely way to explain  the unusual anisotropy in $\kappa/T$, the small size of these linear
terms, and the magnetic field dependence observed.   To some extent our
calculations also depend on the role of disorder in these materials, which is not completely
understood.  In Section II, we present the model we study, in Section III give our results, in Section IV discuss them in comparison
with experiment,  and in Section V, conclude and critically consider the limitations of our approach.


\section{Model}
\label{sec:model}

\subsection{Fermi surfaces and gap structures}
\label{subsec:FS_gap}

The purpose of this study is to examine transport in the superconducting state at very low energies; to this end, only the
structure of the Bogliubov quasiparticle spectrum near gap nodes or deep minima {is} relevant.  In theoretical studies of
these systems performed thus far, nodal structures in $A_{1g}$ states have been obtained on either electron or hole pockets,
but not both. We therefore consider two Fermi surface sheets, one of which ($S_1$) possesses gap nodes. The second ($S_2$)
will be assumed to have deep gap minima, or to be irrelevant altogether for extremely low-energy transport. Results are
symmetrized to ensure invariance under point group rotations. For the 122 systems, we will have in mind primarily that
sheet $S_1$ is one of the hole sheets\cite{s_graser_10} and that $S_2$ is the (properly symmetrized) electron sheet.
The reason is that, to the extent that   nodes are occasionally found in  spin fluctuation calculations on the electron sheets,
they tend to  run vertically from bottom to top of the Brillouin zone, and would therefore lead to an extremely large thermal
conductivity anisotropy $\kappa_{ab}/\kappa_c$, in apparent contradiction to experiment.   However, none of the results actually
depend on the assignment of $S_1$ and $S_2$.

%

\begin{figure}
\includegraphics[width=1.0\columnwidth]{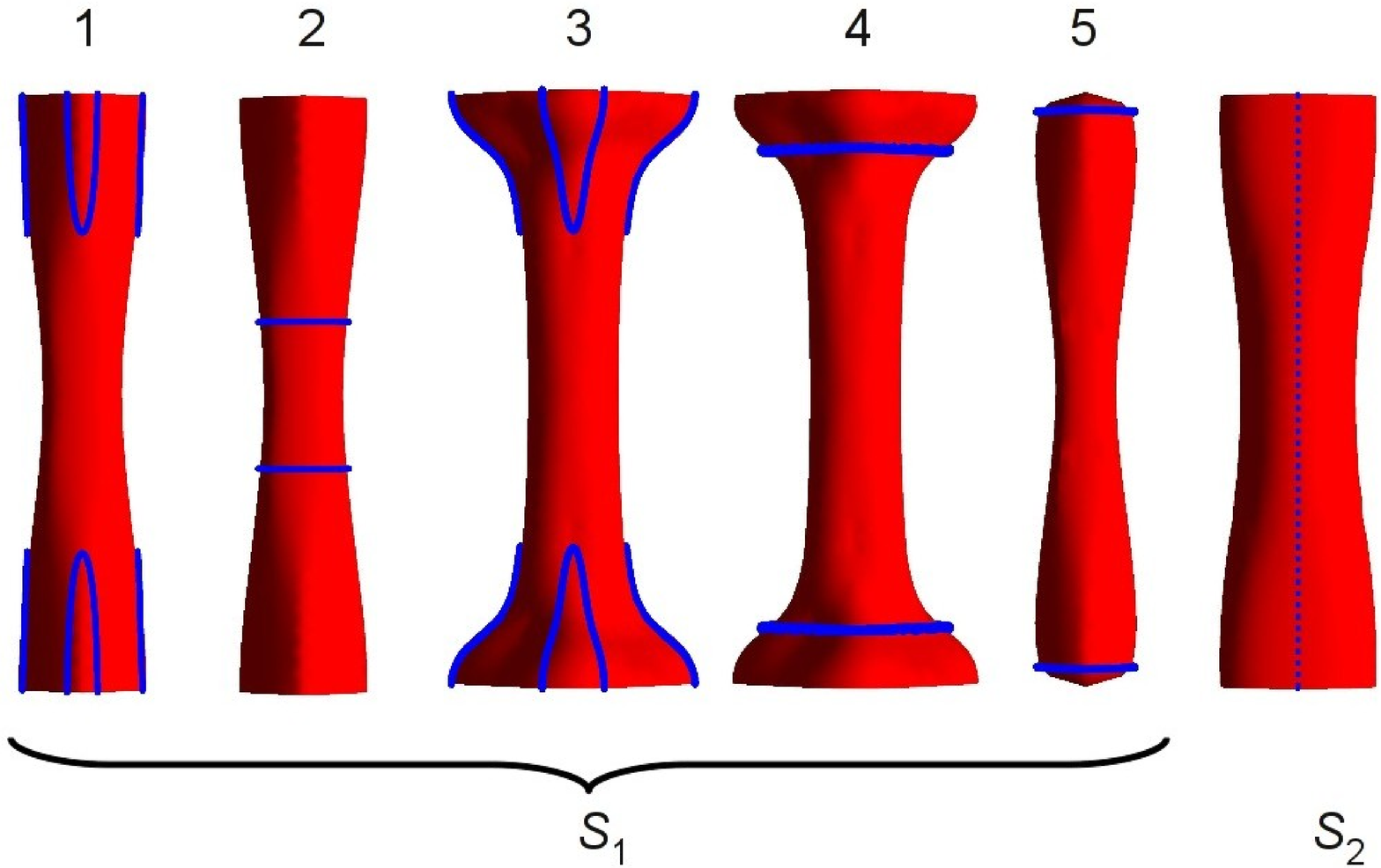}
\includegraphics[width=\columnwidth]{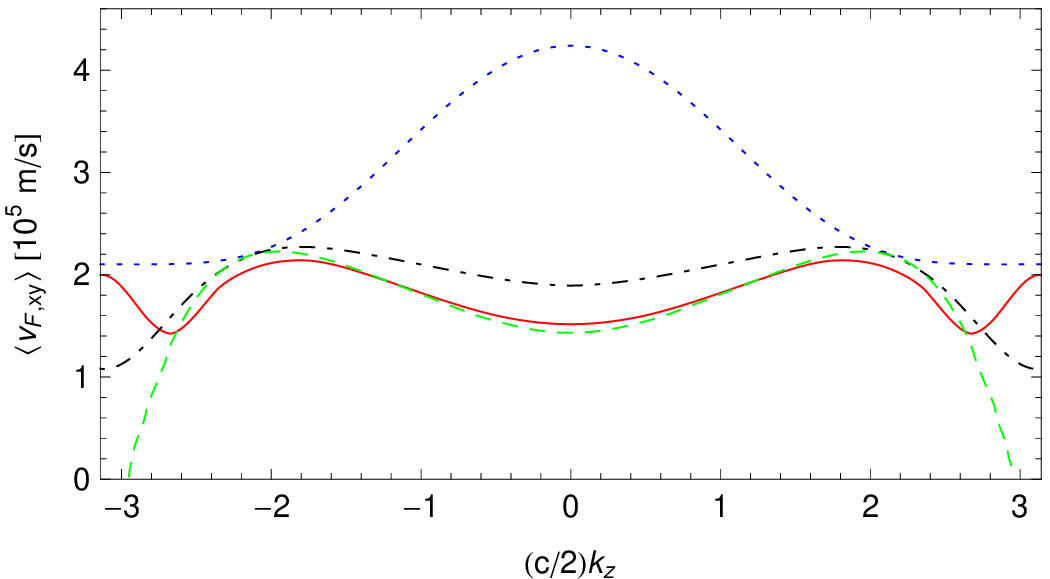}
\includegraphics[width=\columnwidth]{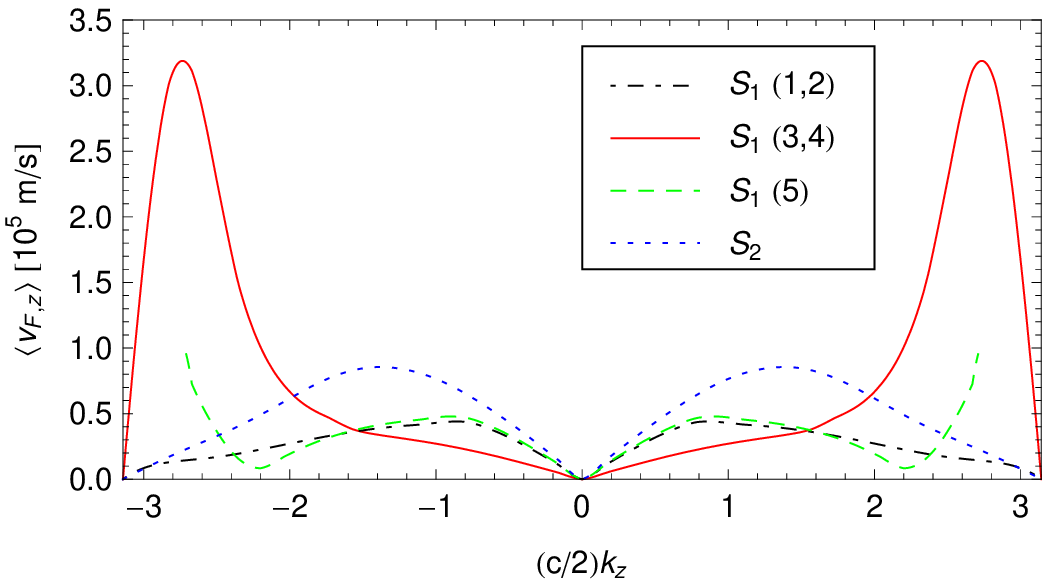}
\caption{Top: various different  Fermi surfaces $S_1$ cases 1-5 and associated gap nodal structures  (dark blue lines) considered in this work.
Top right: sheet $S_2$ considered for all cases, with lines of deep gap minima (dark blue dashed lines) indicated. Bottom: Fermi velocity
components $v_{F,xy}\equiv \sqrt{v_{F,x}^2+v_{F,y}^2}$ and $v_{F,z}$ plotted vs. $k_z$ for the five  $S_1$ cases. Fermi velocities used in calculations
are a constant factor of 2 smaller than those shown here. See text for discussion. }
 \label{fig:FS}
\end{figure}

Fig. \ref{fig:FS} shows various  different kinds
of Fermi surfaces $S_1$ and nodal structures which we consider in this work.
Among the figures labeled ``$S_1$", example 1 is a Fermi surface fit to the hole-doped $\alpha_1 $   sheet of Ref. \onlinecite{s_graser_10},
calculated using a 5-orbital fit to a first principles calculation for \BFA~using the Quantum ESPRESSO package, assuming fixed experimentally
determined atomic positions in the unit cell.
The nodal line structures considered here are typical of the 3D ground
states found on the
$\alpha_1$ and $\alpha_2$ sheets in that work. Example 2 is the same
Fermi surface with horizontal nodes found in the same case in Ref.
\onlinecite{s_graser_10}
on the $\alpha_1$ sheet.  Example 3 is a model Fermi surface with
identical topology but increased flaring near the top and bottom zone
faces,
with assumed V-shaped nodes near the flared portion. This Fermi surface
is obtained within the same calculational scheme\cite{s_graser_10} for
the $\alpha_2$ Fermi surface,
but using a structural optimization of the internal coordinate of the As
position. In both calculations we have used $a=3.9625$~\AA  ~and
$c=13.0168$~\AA.  Similar results are found
elsewhere\cite{Singh2008,Sefat2008,Wang2010a}.
ARPES has provided some evidence for flaring of  hole Fermi surfaces
with large $v_{F,c}$  near the zone top
\cite{Malaeb2009,Fink2009a,Thirupathaiah2010,Kondo2010}, increasing with
Co concentration, but at this writing there is no complete
consensus on the 3D Fermi surface of the doped materials. Example 4 is
the same Fermi surface but with horizontal nodes close to the $k_z$
value where $v_{F,c}$ takes on a maximum.  Example 5 is a model surface
similar to that found in density functional theory calculations upon
electron doping of the $\alpha_1$ sheet  together with a model gap
structure giving nodes on the top surface.
 Finally, the last panel in
Fig.~\ref{fig:FS} shows the second sheet $S_2$, fit to the Quantum ESPRESSO calculation for the electron sheets of \BFA~for unrelaxed As
coordinate, which is assumed to exist together with each of the cases 1-5 for $S_1$. We emphasize that our point of view here is that of
phenomenologists; we allow ourselves to ask what would be the consequences if the Fermi surfaces actually took the forms shown for any
of the doped materials.

A crucial ingredient in the calculations presented below are the  Fermi
velocities near the nodal surfaces. In Table \ref{Fermi_velocities}, we list some values of Fermi velocity components and certain
averages which will be important in the discussion below for the various cases for sheet $S_1$ and our fixed choice of $S_2$.
The $k_z$ dependence of the Fermi velocity components is exhibited in Figure \ref{fig:FS}.

\begin{table}
\caption{{The DFT root mean square (rms) values of the in-plane
and the out-of-plane component of the Fermi velocity on the different
hole pockets $S_1$ and on the electron pocket $S_2$.
In addition the rms values of the Fermi velocity components
on the hole pockets, averaged only over the nodal regions,
are tabulated. All values have been renormalized downward by a factor of 4 to account many-body effects and, are given in units of $10^5 \frac{m}{s}$.  In the last lines are listed the
$r_z$ parameters in Eqs. \ref{eq:OP1}-\ref{eq:OP2} which determine the order parameter on the $S_1$ Fermi surfaces used to calculate results in this work,
and the gap magnitude parameter $\Delta_0$(meV).}}
\label{Fermi_velocities}

\renewcommand{\arraystretch}{1.4}
\begin{tabular*}{\columnwidth}{@{\extracolsep{\fill}}ccccccc}
\hline \hline
 &$S_1 (1)$&$S_1 (2)$&$S_1 (3)$&$S_1 (4)$&$S_1 (5)$&$S_2$\\\hline
$v_{F\perp}$ & 1.977 & 1.977 & 1.826 & 1.826 & 1.74 & 3.021 \\
$v_{F,z}$ & 0.293& 0.293 & 1.277 & 1.277 & 0.683& 0.562 \\
$v_{F\perp}^\mathrm{node}$ & 2.074 & 2.037 & 2.076 & 1.546 & 1.067 &   \\
$v_{F,z}^\mathrm{node}$ & 0.473& 0.433 & 1.197 & 2.591 & 0.792 &   \\
$r_z$ & 0.9 & -1.4 & 0.9 & 1.2 & 1.1  & 0.9 \\
$\Delta_0$ & -8.6 & 9.1 & -8.6 & -9.4 & -8.4 & 1.5\\
\hline \hline
\end{tabular*}
\end{table}

%
%

In calculations of observables presented below, we have {\it reduced} all Fermi
velocities shown in Table 1 by a factor of 4 to account for the effective renormalization of
the bands with respect to DFT seen in ARPES which appears to be between 1.5 and 6 for
all Fe-based superconducting materials\cite{Fink2009a,Brouet2009}.  There may be an additional low-energy renormalization at the
meV scale, as discussed by Benfatto \etal\cite{Ortenzi2009}.  All of these many body effects are lumped
into a single doping-- and momentum--independent renormalization of the Fermi velocity here, to get the
crudest description of the anisotropy of transport properties driven by the underlying anisotropy
of the DFT Fermi surface.

The model we adopt for the order parameter is a 3D extension of the phenomenological form treated, e.g. in Ref. \onlinecite{v_mishra_09}.
To obtain strictly horizontal nodes (cases 2,4,5), we consider gaps of the form
\begin{equation}
\Delta (k_z) = \Delta_{0} [1+ r_{z} \cos(k_{z} d)],
 \label{eq:OP1}
\end{equation}
whereas V-shaped nodal structures (cases 1,3) similar to that found in Ref. \onlinecite{s_graser_10} (see Fig. \ref{fig:FS}) are produced by
\begin{equation}
\Delta (k_z , \phi) = \Delta_{0} [1+ r_{z} \cos(4\phi) (1-\cos(k_z d))],
 \label{eq:OP2}
\end{equation}
with $r_z>1$ for the first kind of order parameter and $r_z>0.5$ for the second kind of order
parameter. In Fig. \ref{fig:FS} for simplicity we have shown only the nodal surfaces of the order parameters chosen on
 the various Fermi surfaces, together with the variation of the Fermi velocity on those surfaces as a function of $k_z$.

\subsection{Disorder in 2-sheet pairing model}
\label{subsec:2sheet}

For disorder, we will assume an orbital-independent matrix element
which scatters quasiparticles either within a given band with
amplitude $U_{ii}$, $i=1,2$, or between bands with amplitude
$U_{12}$.  We sum all single-site
scattering processes of arbitrary strength to obtain a
disorder-averaged Nambu self energy ${\underline\Sigma} = n_{imp}\underline{T}$
\,, where $n_{imp}$ is the concentration of
impurities, and ${\underline{T}}$ is the impurity $T$-matrix
 as parameterized e.g. in Ref.~\onlinecite{v_mishra_09}. For simplicity, we assume $U_{11}=U_{22}\equiv U_d$,
with equal densities of states $N_i=N_0$ throughout the paper. In
our preliminary considerations we restrict ourselves to purely
intraband scattering, $U_{12}=0$. The disorder is
characterized by two intraband scattering parameters on each
sheet: $\Gamma_i\equiv n_{imp}/(\pi N_i)$ and $c_i=1/(\pi
N_iU_{ii})$; For our simple initial case with 2 symmetric bands we
set $\Gamma_{i}=\Gamma$ and $c_i=c$, $i=1,2$.   We only consider intraband scattering in
the unitary limit \cite{a_kemper_09}, and further note that nonzero interband scattering
does not affect the physics qualitatively, unless the interband
scattering is as strong as the intraband scattering\cite{Mishra2009}.
In presence of such strong isotropic scattering, one would expect
a large suppression of $T_c$ which is
not observed in experiments.

The disorder-averaged matrix Green's function in the presence of scattering in
the superconducting state is given  by a diagonal matrix in
band space,
\begin{equation}
{\underline G}({\bf k}, \omega) = {\tilde\omega\tau_0
+\tilde\epsilon_\k\tau_3+\tilde\Delta_{\bf k}\tau_1\over
\tilde\omega^2-\tilde\epsilon_\k^2-\tilde\Delta_\k^2},
\label{Greensfctn}
\end{equation}
where $\k = \k_i\in S_i$ is restricted to Fermi surface sheet
$S_i$ with $i=1,2$, and the renormalized quantities
$\tilde\omega\equiv \omega-\Sigma_0$, $\tilde\epsilon_\k\equiv
\epsilon_\k+\Sigma_3$, $\tilde \Delta_\k \equiv
\Delta_\k+\Sigma_1$ also depend on the band indices through $\k$.
The $\Sigma_\alpha$ are the components of the self-energy
proportional to the Pauli matrices
$\tau_\alpha$ in particle-hole (Nambu) space.

\section{Results}
\label{subsec:dos_kappa_lambda}

\subsection{DOS}

\begin{figure}
\includegraphics[width=1.0\columnwidth]{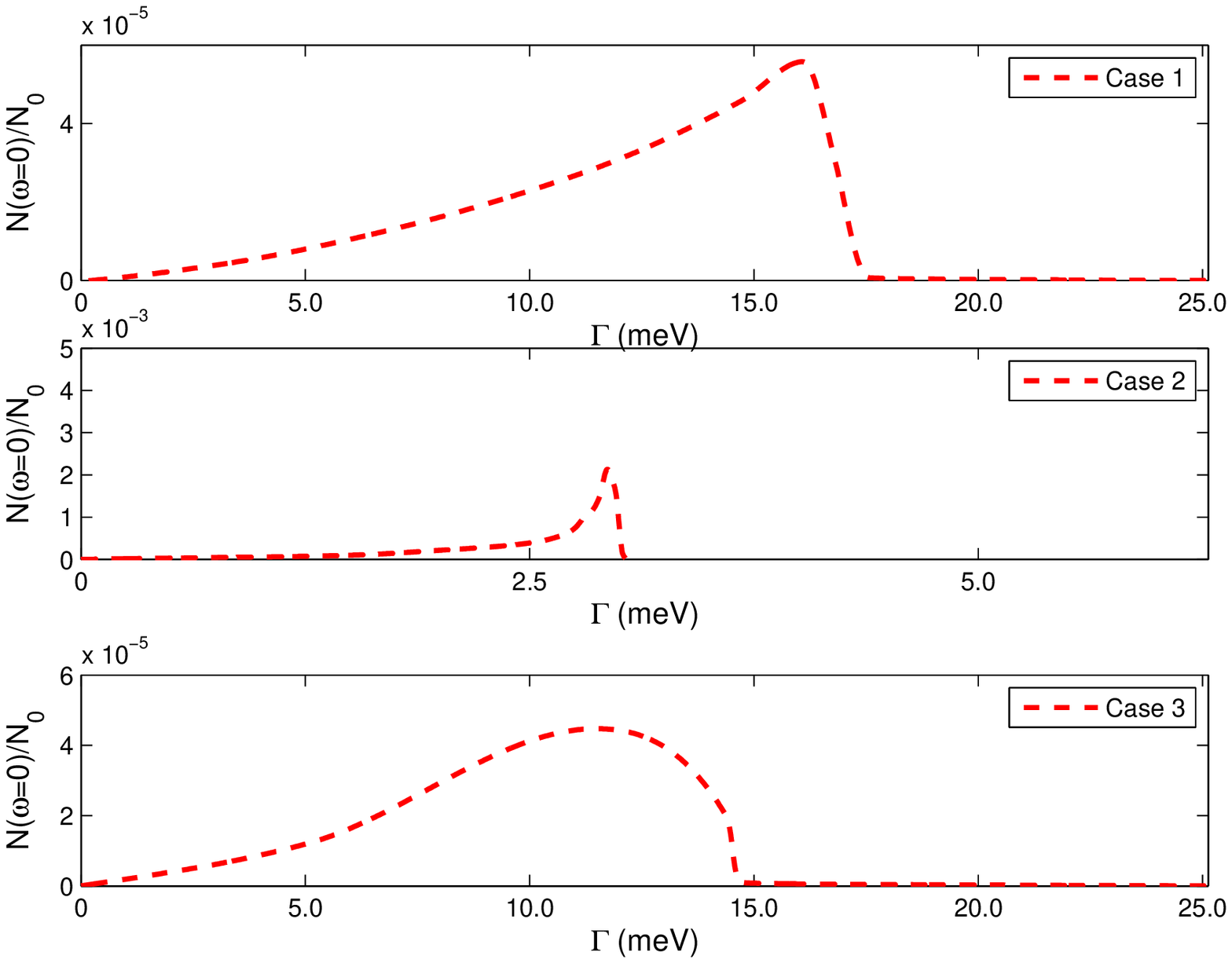}
\includegraphics[width=1.0\columnwidth]{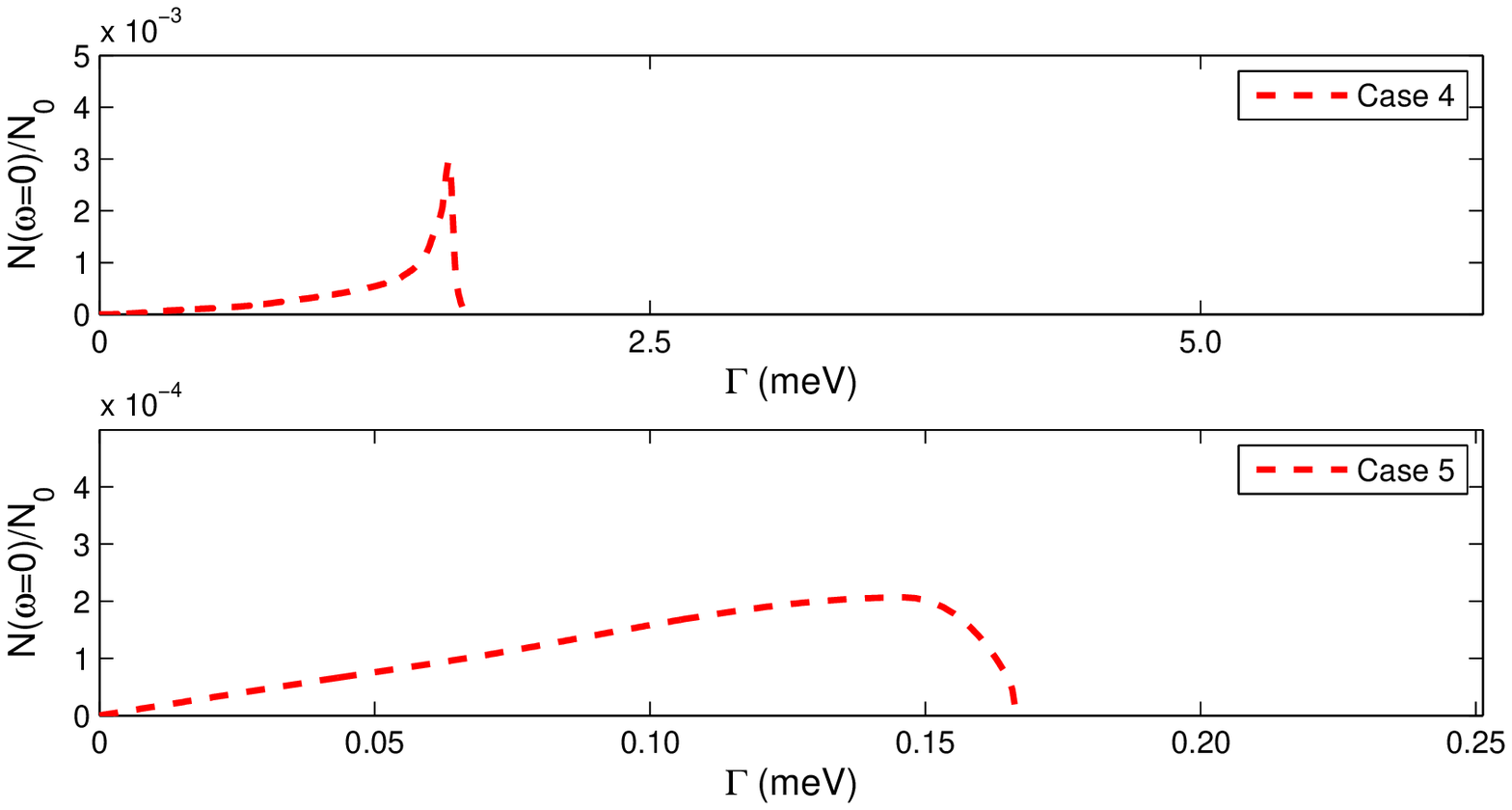}
\caption{Residual density of states $N(\omega=0)$ normalized to the total normal state density of state at the Fermi
energy $N_0$ vs. total unitary intraband scattering rate parameter $\Gamma$ in meV. Figures (a)-(e) correspond to cases 1 to 5, respectively. }
 \label{fig:DOS}
\end{figure}

{Below we focus on several bulk observables. It is useful to start with the
analysis of the total density of quasiparticle states (DOS)}
\begin{equation} 
N(\omega) = -{1\over 2\pi} {\rm Tr~}{\rm Im} \sum_{\k_{i}} {\underline G}(\k_{i},\omega),
\end{equation}
{where the momentum summation indicates the explicit integration over
distinct Fermi surface sheets with momenta $\k_i$.} The total residual density of states (DOS)
at the Fermi level $N(0)$  contains vital information about the low lying quasiparticle excitations and will
determine all leading low temperature power laws. This
quantity is  sensitive to disorder  or
magnetic field. To give the reader a sense of the average density of quasiparticle
excitations without reference to Fermi surface anisotropy, we calculate the residual DOS as a function of
normal state scattering rate due to intraband disorder.  Figure  \ref{fig:DOS} shows
 $N(0)$ for the five different cases in consideration; generally it first increases
with disorder, attains a maximum value and rapidly drops to zero again.  This ``re-entrant"
gapped behavior is a consequence of the node lifting phenomena characteristic of accidental
nodes in an order parameter with A$_{1g}$ symmetry\cite{l_borkowski_94,v_mishra_09}.  Considering
the evidence provided by Refs. \onlinecite{c_martin_10,Reid2010} for low-energy excitations,
we assume in our discussion below that we are in the regime with nodes and no true
gap in the system.  The values of the scattering rate $\Gamma$ for ``dirty" systems presented here
are chosen in each case to correspond to the maximum residual DOS induced by disorder as
shown in Fig. \ref{fig:DOS}.  In all cases we have verified that the corresponding  $T_c$ suppression relative to the pure system is
 ${\cal{O}}$(1\%) or less  for
all cases.

We note that for the various cases considered, the node-lifting process can take place at quite
different rates.  In particular, the nodes on the top of the closed Fermi sheet in case 5 are
quite sensitive to a very small amount of disorder.  The sensitivity of this case to disorder is
the strongest argument against an explanation of the Reid \etal~data based on a Fermi surface and state of  type 5.

\subsection{Penetration depth}

The London penetration depth $\lambda_i$ for currents flowing in the $i$th direction is a fundamental measure of the superconductivity and
is related to the superfluid density by  $\rho_s/m^*= (\mu_0e^2\lambda^2(T))^{-1}$. The superfluid density tensor
is related to the total electromagnetic response of the system to an external field in the $\q\rightarrow 0, \omega=0$
limit. For a general multiband dispersion $\epsilon_i(\k)$ and in presence of the impurities, one finds that the penetration
depth is given by the expression

\begin{eqnarray}\label{eq:pendepth}
{1\over\lambda_{\alpha}^2} &=&{2\mu_0 e^2\over d\hbar^2} \sum_{i} m_{i} \int_{0}^{\infty} \frac{d\omega}{\pi} \left\langle
\tanh\left(\frac{\omega}{2T}\right) \right. \\ \nonumber
&\times& \left[ ({\bf v_{F,i}.\hat{\alpha}})^{2} {\rm Im} \left(\frac{\tilde{\Delta}_{i}^{2}} { (\tilde{\Delta}_{i}^{2}
-\tilde{\omega}_{i}^{2})^{\frac{3}{2}}} \right)  \right. \\ \nonumber
&-& \left. \left. 2({\bf v_{F,i}.\hat{\alpha} v_{\Delta,i}.\hat{\alpha} }) {\rm Im }\left( \frac{\tilde{\Delta}_{i}
\tilde{\omega}_{i}}{(\tilde{\Delta}_{i}^{2}-\tilde{\omega}_{i}^{2})^{3/2}} \right) \right] \right\rangle_{\phi,\k_z}
\end{eqnarray}
where $d$ is the distance between planes, the index $i$ denotes the band and $\alpha$ stands for the current direction $ab$ or $c$
and $\langle \dots \rangle_{\phi,k_z}$ is an average over the Fermi surface. In the absence of disorder, this expression reduces to the result for
the superfluid density of a clean system with general band found, e.g. in Ref.~\onlinecite{Sheehy2004}.  The second term in
Eq.~(\ref{eq:pendepth}) contains the nodal gap slope $v_{\Delta}\equiv \partial \Delta_\k/\partial \k$, which is extremely small
compared to the Fermi velocity, and will be neglected in our calculations.  Most experiments measure the change in the penetration
depth $\Delta\lambda$  from some minimum temperature $T_{min}$, and are not sensitive to the absolute value of the penetration depth.

\begin{table}
\caption{Zero temperature penetration depth. }
 \begin{tabular}{ccc}
  \hline
 Case & $\lambda_{0,ab}$ (nm) & $\lambda_{0,c}$ (nm) \\
\hline
\hline
1,2 & 87 & 334 \\
3,4 & 87 & 287 \\
5 & 87 & 329 \\
\hline
\hline
 \end{tabular}
 \label{table:lam0}
\end{table}

Note that since we begin from a realistic Fermi surface and assume a $T_c\sim 25K$ as appropriate for the Co-doped
122 systems, the reader might assume that we are in a position to calculate on an absolute length scale the zero
temperature penetration depth $\lambda_0$. Indeed the expression (\ref{eq:pendepth}) reduces at $T=0$ to
 \begin{eqnarray}
 \frac{1}{\lambda_0^2} = \frac{2 \mu_0 e^2   }{ \hbar^2 d \pi}\sum_{i}   \left\langle m_i^{*} v_{F,i}^{2}\right\rangle_{FS}\label{lam_zeroT}
\end{eqnarray}
which gives the correct result in the limit of  a single parabolic band. However a closer
examination of Eq.~(\ref{eq:pendepth}) reveals that the integrand determining $\lambda_0$
extends over all occupied bands, i.e. it involves the full electronic structure and not just
the Fermi velocities, whereas in Eq.~(\ref{eq:pendepth}) we have linearized the band structure
near the Fermi level for simplicity. Thus we are not able to use Eq.~(\ref{eq:pendepth}) to
obtain accurate estimates of $\lambda_0$ even if the input band structure $\epsilon_i(\k)$ was perfectly correct.
By contrast, the integrand determining
\begin{equation}
  {\Delta \lambda_\alpha(T) } \simeq {\lambda_{0,\alpha}\over 2} \left[ {\left(\lambda_{0,\alpha}\over \lambda_{\alpha}\right)^2}-1\right]
\end{equation}
is sharply peaked at the Fermi level, and so may be calculated accurately. Even in this case,
however, the overall scale is set by $\lambda_{0,\alpha}$. The usual procedure would be to take
$\lambda_{0,\alpha}$ from optical experiments.  Indeed, in these systems the $ab$-plane penetration
depth is of order 200-450 nm \cite{Williams2009,Gordon2009,Gordon2010,Luan2010,Luan2010a,Nakajima2010}, but there are no
reported measurements of $\lambda_{0,c}$ of which we are aware.  For purposes of this work
we use $\lambda_0$'s determined from Eq.~(\ref{lam_zeroT}) (given for completeness in table \ref{table:lam0}) to fix the scale
of $\Delta\lambda_\alpha$. The overall length scale of field penetration may therefore be incorrect by up to a factor of two,
but the $T${\it-dependence} should be calculated accurately.

\begin{figure}
\hspace*{-.5cm}
\includegraphics[width=1.0\columnwidth]{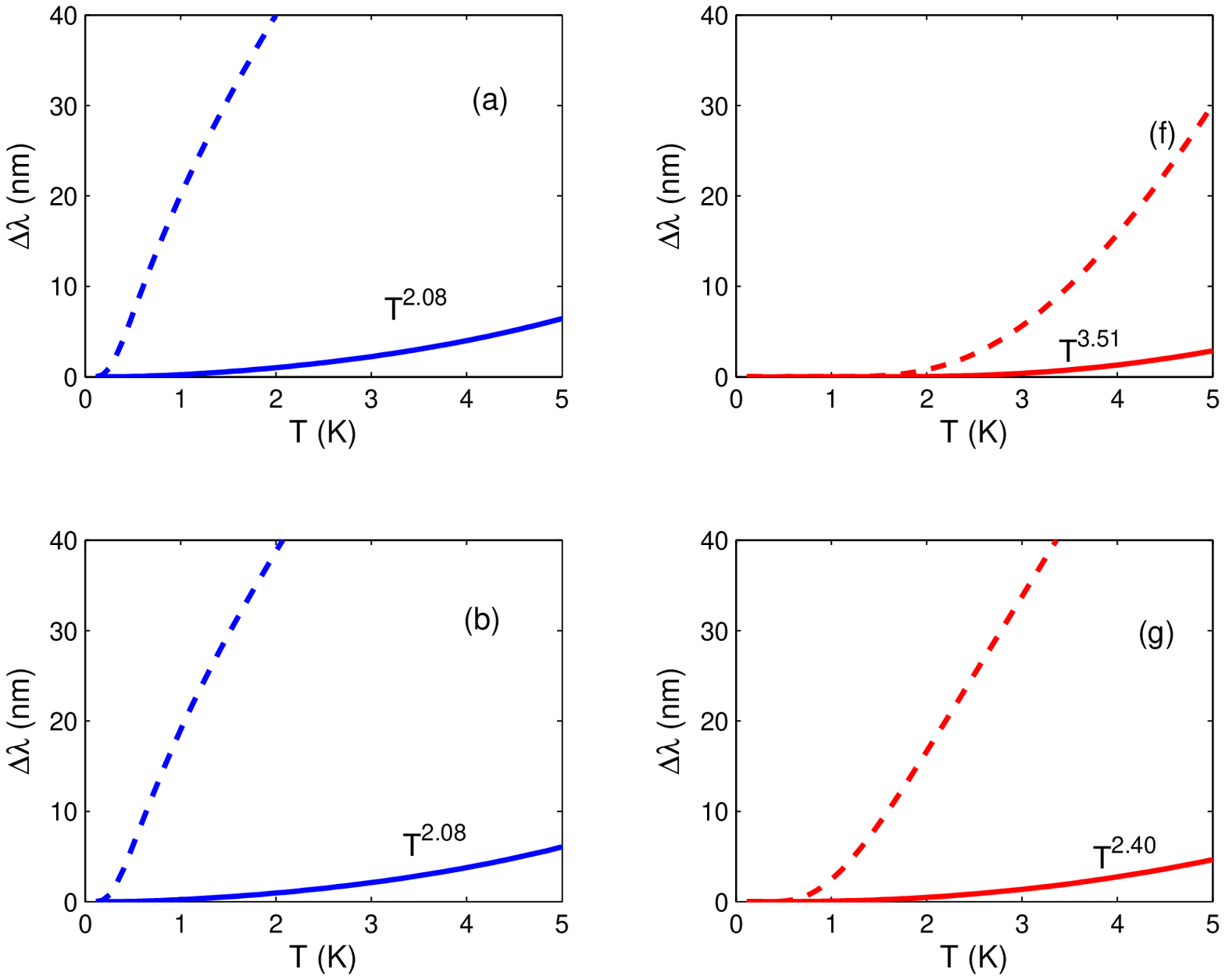}
\hspace*{-.5cm}
\includegraphics[width=1.0\columnwidth]{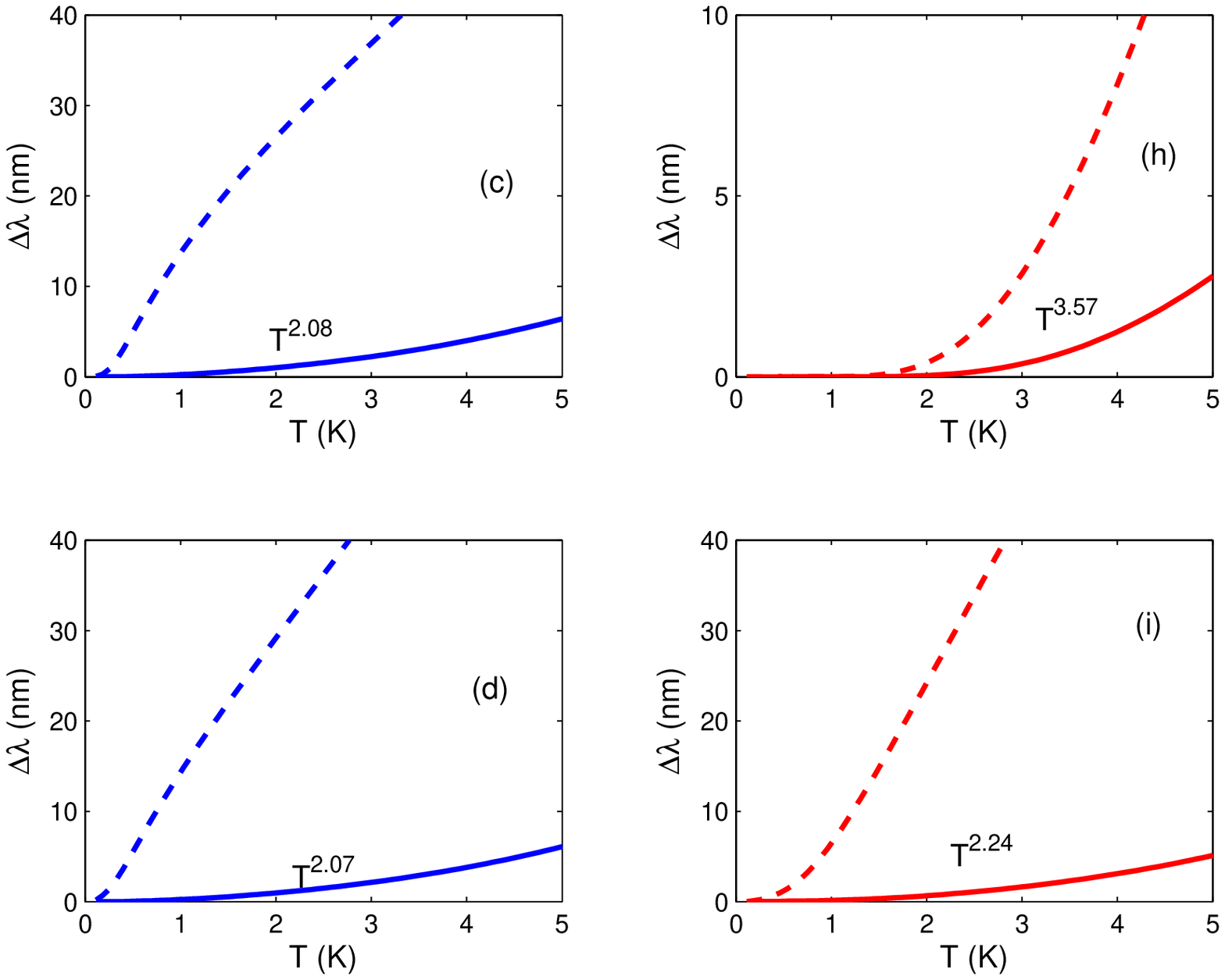}
\hspace*{-.5cm}
\includegraphics[width=1.0\columnwidth]{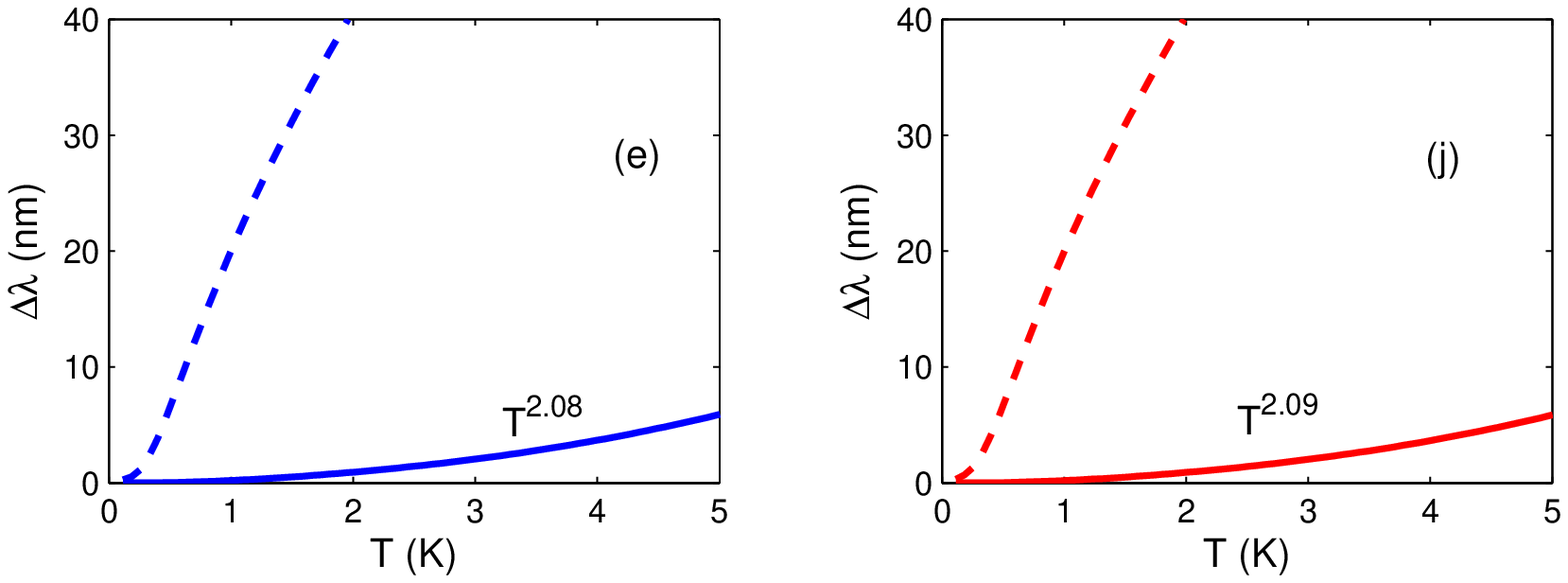}
\caption{ Change in penetration depth $\Delta\lambda$ measured in nm vs. {temperature  $T$}.
Panels (a)-(e): clean penetration depth for $c$-axis (dashed lines) and $ab$-plane (solid lines) currents for cases 1-5.
(f)-(j): same plots for  dirty system with $\Gamma$ chosen in each case to give the maximum
$N(\omega=0)$ in Fig. \ref{fig:DOS}. For $\lambda_{ab}$, power laws over the $T$ range shown in the panel are indicated. }
 \label{fig:pened}
\end{figure}

The temperature correction $\Delta \lambda(T)$ is often used as a probe of the nodal structure of gaps in unconventional
superconductors\cite{Prozorov2006}. In general, the asymptotic $T \rightarrow 0$ power
laws for systems with line nodes are $\Delta\lambda\sim T$ for a clean and $\Delta \lambda \sim T^2$ for a dirty system
\cite{Einzel1986}, but the range of validity of these results may be experimentally unobservable depending on the form of
the impurity scattering and the detailed momentum dependence of the superconducting gap.
In our current calculations with intraband impurity scattering only,
the concentrations have been chosen in order to minimize $T_c$ suppression; the corresponding density of states and size of
the asymptotic $T^2$ terms are therefore quite small, and the latter is  not clearly visible in Fig. \ref{fig:pened}.
On the scale of the Figure shown, these asymptotic power laws are modified by low energy scales associated with both sheets $S_1$
and $S_2$, as well as the significant energy dependence of the impurity self-energy. Over an experimentally relevant $T$ range,
a wide range of ``best fit" power laws are therefore a priori possible.

In Fig. \ref{fig:pened}, we show $\Delta\lambda$ as a function of temperature for the various cases we
consider here. The size of the observable quantity $\Delta\lambda$ is seen to be in reasonable agreement
with the measured penetration depth changes in recent measurements in optimal to overdoped Ba-122\cite{c_martin_09}.
The scale of $\Delta\lambda_c$ is significantly larger than  $\Delta\lambda_{ab}$, due to the much larger $\lambda_{0,c}$.
In addition, Figure~\ref{fig:pened} shows qualitatively different behavior for $ab$-plane and $c$-axis penetration depths.
These results, particularly those for cases 3 and 4, are in qualitative  agreement with experimental measurements on Ni
doped 122 compounds, where  linear $T$ behavior along the $c$-axis and power law behavior for $ab$-plane with exponent
$~1.6-2.5$ for concentrations $x$=0.03 to 0.07 (only those above $x\sim 0.04$ are outside the spin density wave (SDW) state)\cite{c_martin_10}
were reported. For Co doped 122 systems, we are not aware of any $c$-axis penetration depth measurements.

The $ab$-plane penetration depth data in 122 systems has been reported for some time now
to be close to a power law $T^\alpha$ with an exponent $\alpha$ of about 2.
It has usually been assumed that this result is obtained due to impurity
scattering, either in a nodal system\cite{v_mishra_09}
or in a fully gapped $s_\pm$ state with interband scattering such that midgap bound states are formed\cite{a_vorontsov_09}.
These effects may indeed be present in very dirty samples.  However, the existence of a linear-$T$ term for the $c$-axis penetration depth
suggests that this simple argument cannot be the complete explanation for  the  approximate $T^2$ behavior in the better samples;
were the system sufficiently dirty to show $T^2$ behavior in the $ab$-plane, it would perforce manifest the same behavior for currents in the
$c$-axis direction (we ignore for the moment the possibility that impurity scattering is itself highly anisotropic with respect to the $ab/c$
directions.).  The difference in power laws in the two directions must therefore be ascribed to some other effect or combination of effects.
In our proposal, there is no true $T^2$ behavior except perhaps at the very lowest temperatures; the upward curvature observed over a range
of several Kelvin is due in part to the ``activation" of a new source of quasiparticle excitations as the temperature is increased.
At low temperature all quasiparticles come from sheet $S_1$, since we have assumed that this is the only sheet where true nodal excitations exist.
On the other hand the order parameter on the $S_2$ sheet is assumed to have deep minima $\Delta_{min}$.  When $T$ exceeds this scale,
quasiparticles from this sheet may be excited and a new contribution (upward curvature) to $\Delta\lambda$ appears.  Since $S_2$ is assumed
to have smaller $v_{F,c}$, it does not contribute significantly to $\Delta\lambda_c$, however.
The second cause of ``power law" behavior is indeed the disorder, but it is impossible to make universal statements about
the large $T$ ranges over which experiments are typically fit.  To give a more useful measure of the form of these results in Fig. \ref{fig:pened},
we have performed best fits of the $T$ dependence to $\Delta\lambda \sim T^\alpha$ over the finite range of temperatures shown.

\begin{figure}
\includegraphics[width=1.\columnwidth]{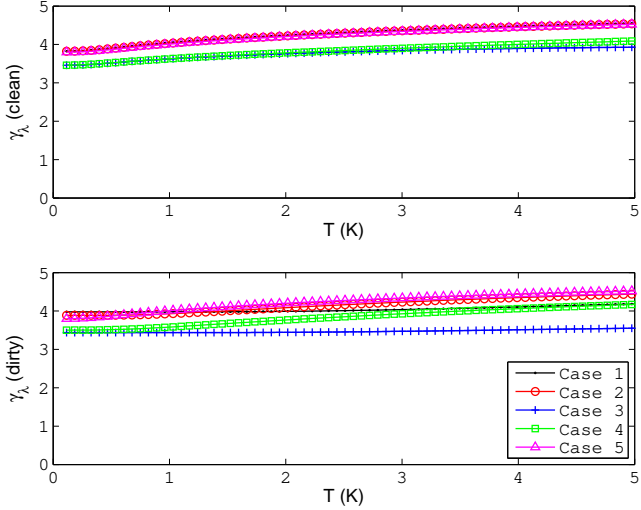}
\caption{Penetration depth anisotropy $\gamma_\lambda(=\lambda_c / \lambda_{ab})$ as a
function of temperature. Top: clean systems  with $\Gamma$ =0.0.  Bottom: dirty systems with $\Gamma$ chosen as
in Fig. \ref{fig:pened}.}
\label{fig:gamma_lambda}
\end{figure}

The  penetration depth anisotropy
\begin{equation}
 \gamma_\lambda\equiv {\lambda_c\over \lambda_{ab}}\simeq {\lambda_{0,c}\over\lambda_{0,ab}}\left(1+{\Delta\lambda_c\over \lambda_{0,c}}
-{\Delta\lambda_{ab}\over \lambda_{0,ab}}\right)\label{pendepth_anisotropy}
\end{equation}
has been claimed in Refs.~\onlinecite{Tanatar2008, Prozorov2009} to be of order 3-6 at low temperature, and to decrease weakly with increasing
temperature, but cannot yet be measured directly.  Experimentally, $\Delta \lambda_i$ is measured, and must be combined with measurements
of $\lambda_{0,i}$, with $i=c,ab$ to determine the anisotropy ratio.  Reliable measurements of $\lambda_{0,c}$ from, e.g. optics experiments, are
not yet available for the 122 materials to our knowledge.  However, within our theoretical framework $\gamma_\lambda$ can be
determined, with the caveats described above, and Fig. \ref{fig:gamma_lambda}
shows $\gamma_\lambda$ for the different cases.  We therefore regard this as a prediction of weakly increasing $\gamma_\lambda$ with temperature
which may be verified at the point that reliable values of $\lambda_{0,c}$ are obtained from optics or elsewhere.
Our results are in the range of anisotropy ratios deduced from experiment\cite{Tanatar2008}, but
increase weakly rather than decrease weakly with temperature
at low temperatures.  A glance at the approximate inequality in Eq.~(\ref{pendepth_anisotropy})  shows that a somewhat smaller value of
$\lambda_{c,0}$ than effectively assumed in Refs. \onlinecite{Tanatar2008, Prozorov2009} would bring the measured $\Delta \lambda_i$'s
and therefore the $T$-dependence of $\gamma_\lambda$ in agreement with our results.

We can now ask which type of Fermi surface/order parameter combination represents the  experimental penetration depth
data best.  For cases 1 and 2, the $c$-axis Fermi velocity is small, which means not only that $\Delta\lambda_c$ is smaller
at low $T$, where it derives entirely from the $S_1$ nodes, but also that the contribution from the $S_2$ deep minima is
correspondingly more important; as seen in Fig. \ref{fig:pened}, for these cases the $c$-axis penetration depth as a function
of $T$ has substantial upward curvature, which is never observed in experiment.  Power law fits to the $ab$-plane
results give exponents which tend to be too close to 1.  On the other hand for cases 3 and 4, $\Delta \lambda_c(T)$ has no upward
curvature above a scale corresponding to approximately $1K$ and is quasilinear over a significant $T$ range, while $\Delta \lambda_{ab}(T)$
displays power law behavior with $\alpha$ approaching 2 in the dirty cases shown over a finite range, as observed in experiment.
Case 5 displays a large penetration depth anisotropy, but an anomalous $T$ dependence which is due to the extremely small $\Delta_{min}$
for this case which must be chosen to place the nodes in a position with high $v_{F,c}$.  Therefore from the point of view of the penetration
depth temperature dependence and crude anisotropy, only cases 3 and 4 appear compatible with experiment. We emphasize again that the power
laws stated in Fig.~\ref{fig:pened} are phenomenological fits and are not exact in any asymptotic limit.  Exponents may be
expected to change as the low energy gap scales are changed, and in particular to increase somewhat as interband scattering is added.

Before closing this section, we note that nonlocal electrodynamics are unlikely to play a significant role
in these measurements.  This is because the typical energy scale which cuts off  singularities in nodal superconductors is
$E^\alpha_{nloc} \equiv (\xi_\alpha/\lambda_\alpha)\Delta_0$.  For both directions in the 122 samples, this scale is of order a
few tenths of Kelvin or less.  The crossover in clean samples from linear $T$ to $T^2$ from nonlocal effects will therefore only
complicate interpretations at unobservably low temperatures, where disorder will probably dominate in any case.

\subsection{Thermal conductivity in zero  magnetic field}

The electronic thermal conductivity $\kappa$ is calculated here  {using a straightforward
extension  of the standard approach}\cite{Ambegaokar1965,p_hirschfeld_88,Durst2000} to multiband superconductors\cite{v_mishra_09}.
In the presence of disorder, we evaluate the expression arising from the electronic heat current bubble,


\begin{eqnarray}
 \label{eq:kappa1}
\frac{{\kappa_{\alpha}}}{\kappa_n}&=& \frac{3} {4\pi^2 T_c \sum_i \left\langle {m_{i}\bf (  v_{F,i}.\hat{a})^{2}}/\Gamma_i\right\rangle_{\phi,k_z}} \sum_{i} \int_{0}^{\infty} d\omega  \\ \nonumber
&&\left\langle \frac{m_{i}}{\mathrm{Re}[\sqrt{\tilde{\Delta}_{i}^{2}-\tilde{\omega}_{i}^{2}}]} \left(v_{i-}^{2}+\frac{|\tilde{\omega}|_{i}^{2} v_{i+}^{2}
-|\tilde{\Delta}|_{i}^{2 }v_{i-}^{2} } {|\tilde{\omega}_{i}^{2}-\tilde{\Delta}_{i}^{2}|}\right)\right\rangle_{\phi,k_z}  \\ \nonumber
&\times&{\rm sech}^{2} \left(\frac{\omega}{2T} \right) \frac{\omega^2}{T^2},\\ \nonumber
\end{eqnarray}
where
\begin{eqnarray}
\kappa_{n}&=&\frac{\pi}{12}\frac{n k^{2}_{B} T_{c}}{\hbar d} \sum_i \left\langle {m_{i}\bf (  v_{F,i}.\hat{a})^{2}}/\Gamma_i\right\rangle_{\phi,k_z}
\label{eq:kappan}
\end{eqnarray}
and $\alpha=a,c$, $m_{i}$ is the band mass,  ${\bf v}_{F,i}$ is the Fermi velocity on the $i$th band, and $v_{i\pm}^{2}\equiv v_{F,i}^{2}\pm {(\partial_\alpha\Delta_{\k,i})}^{2}$.
%
%
Terms involving the gap slope $v_{\Delta}$ at the nodes are numerically negligible since $v_{\Delta}\ll v_F$ and are omitted in our evaluations.
Here, $\langle \dots \rangle_{\phi,k_z }$ is an average over each Fermi surface sheet, $\sum_{i}$ denotes the sum over the bands, $v_{F,i}$ is the
Fermi velocity on sheet $i$, and $n$ is the number of FeAs planes per unit cell.

We focus first on the  linear term in $\kappa$  in the $T\rightarrow 0 $ limit, the only experimentally unambiguous
signature of purely electronic heat transport by nodal excitations.  In the limit $T\rightarrow 0$, Eq.~(\ref{eq:kappa1}) reduces to
\begin{equation}
\frac{\kappa^{i}}{T}\vert_{T\rightarrow0} \simeq a N_\mathrm{nodes} {k_B^2m^*\over \hbar d }  \left[\frac{v_{F,i}^{2}}{k_Fv_{\Delta,i}}\right]_\mathrm{node},
\label{eq:kappalowT}
\end{equation}
where $a$ is a dimensionless constant, $N_\mathrm{nodes}$ is the number of distinct nodal surfaces, assumed equivalent, and
$m^*$ is the effective mass for motion of quasiparticles in the $ab$-plane.  The result obtains exactly only if $v_{F,i}$ and $v_{\Delta,i}$
are constant over the entire nodal surface, which is unrealistic except, crudely speaking, in a 2D situation, but if the velocities
are replaced by the average expression of Eq.~(\ref{eq:kappalowT}) will give a reasonable result.  The constant $a$ reduces to
$1/6$ in the well-known case of a 2D $d$-wave order parameter, and will be of the same order of magnitude
for a clean nodal $s_\pm$ state if the nodes are vertical and run the length of the Fermi cylinder.
In any  situations where the nodal lines do not span the dimensions of the Fermi sheet, i.e. consist of small circles or segments,
as in all the cases for $S_1$ considered in Fig.~\ref{fig:FS}, the factor $a$ contains the relative nodal phase space and will therefore be $a \ll 1$.
Note that in the $d$-wave and certain $p$-wave cases, the result of Eq.~(\ref{eq:kappalowT})  is also universal, in the sense that the nodal velocities
are unrenormalized by disorder~\cite{m_graf_96,p_hirschfeld_96}, but in the generalized $s_\pm$ case this universality breaks down~\cite{v_mishra_09}.

It is instructive to compare crude estimates of what one might expect for $\lim_{T\rightarrow 0} \kappa_i/T$ with
actual theoretical results and the experimentally measured values.  In Table~\ref{table:estimates} we first show progressively more
accurate estimates of this quantity. First, we remark that a small linear-$T$ term in $\kappa_{ab}$  {\it must} be present if nodes are
present and if ${\bf v}_F$ has any nonzero component in the $ab$-plane.  To be consistent with earlier data and with the recent work of
Reid \etal\cite{Reid2010} in the Co-doped Ba-122 system, this term must be quite small, consistent with zero within the resolution of the
experiment, at least near optimal doping. It is clear from the table that a 2D $d$-wave state or, equivalently, an anisotropic nodal
$s$-wave state with line nodes running the length of the Fermi cylinder (``method T1") would dramatically overestimate the experimental results
for $\kappa/T$ in the ab plane.  The best estimate based on Eq.~(\ref{eq:kappalowT}) (``method T2"), accounting for realistic values of
nodal velocities $v_\Delta$, still gives a value significantly larger than experiment near optimal doping in
the \BFCA~system studied by Reid \etal~as shown in Table~\ref{table:estimates}.
This is at least in part because method T2 also does not account for the small nodal phase space. The T3 column of
Table~\ref{table:estimates} shows the clean limit of the exact numerical evaluation of Eq.~(\ref{eq:kappa1}) for the 5 cases.
These results are now of order the scatter in the data (consistent with zero) of the $ab$-plane linear-$T$ terms in $\kappa$
identified in samples near optimal doping by Reid \etal, which we take to be the resolution of the experiment, about $\sim 1\mu $W/K$^2$cm.

\begin{table}[t]
\caption {Theory columns: comparison of different estimates of the $T\rightarrow 0$ limit of Eq.~(\ref{eq:kappa1}) for currents in
the $ab$-plane, expressed in $\mu$W/K$^2$cm for a system with $T_c$=23K and $d=13$\AA, with Fermi velocities and order parameters
for the various cases as given in Sec.~\ref{subsec:FS_gap}.
Method T1: 2D $d$-wave universal result using $\langle v_{F}\rangle_{rms}$ from Table~\ref{Fermi_velocities} and
$k_F\langle v_{\Delta}\rangle_{rms}\sim 2 k_B T_c$ and $a=1/6,N_\mathrm{nodes}=4$ in Eq.~(\ref{eq:kappalowT});
Method T2: Eq.~(\ref{eq:kappalowT}) using $\langle v_{F,a}\rangle_{rms}$ and $\langle v_{\Delta}\rangle_{rms}$ averaged over the
nodal manifold for each case. Method T3: Full numerical result from Eq.~(\ref{eq:kappa1}).
\BFCA: Experimental data for asymptotic low-$T$ linear-$T$ term in $\kappa_{ab}$ from Ref.~\onlinecite{Reid2010} for
values of $x$ shown. Numbers in parentheses are values of extrapolated linear term in $\mu$W/K$^2$cm for different samples;
zero is reported if these are negative. LFPO: value for LaFePO from Ref.~\onlinecite{Yamashita2009};
P-122 value for \BFAP with $x$=0.33 from Ref.~\onlinecite{Hashimoto2010}}
\begin{center}
\begin{tabular}{cccccccccc}\hline
\multicolumn{4}{c}{Theory}&\multicolumn{4}{c}{\BFCA }&\multicolumn{1}{c}{{LFPO}}&\multicolumn{1}{c}{P-122}  \\
\hline
Case & T1 & T2 & T3 &0.074&0.108&0.114&0.127& &$0.33$\\
\hline
1 & 118  & 31  &  1.8 &&&&&&  \\
2 & 118   & 71 & 3.1 &&&&&&\\
3 & 102  & 31  & 2.5 &0&0&0&17&{3000}&250 \\
4 & 102 & 59  & 6.3 &(-1,3)&(-1)&(-9,-13)&&& \\
5 & 94  & 47  & 3.1 &&&&&& \\
\hline
\end{tabular}
\end{center}\label{table:estimates}
\end{table}

\begin{figure}
\includegraphics[width=1.0\columnwidth]{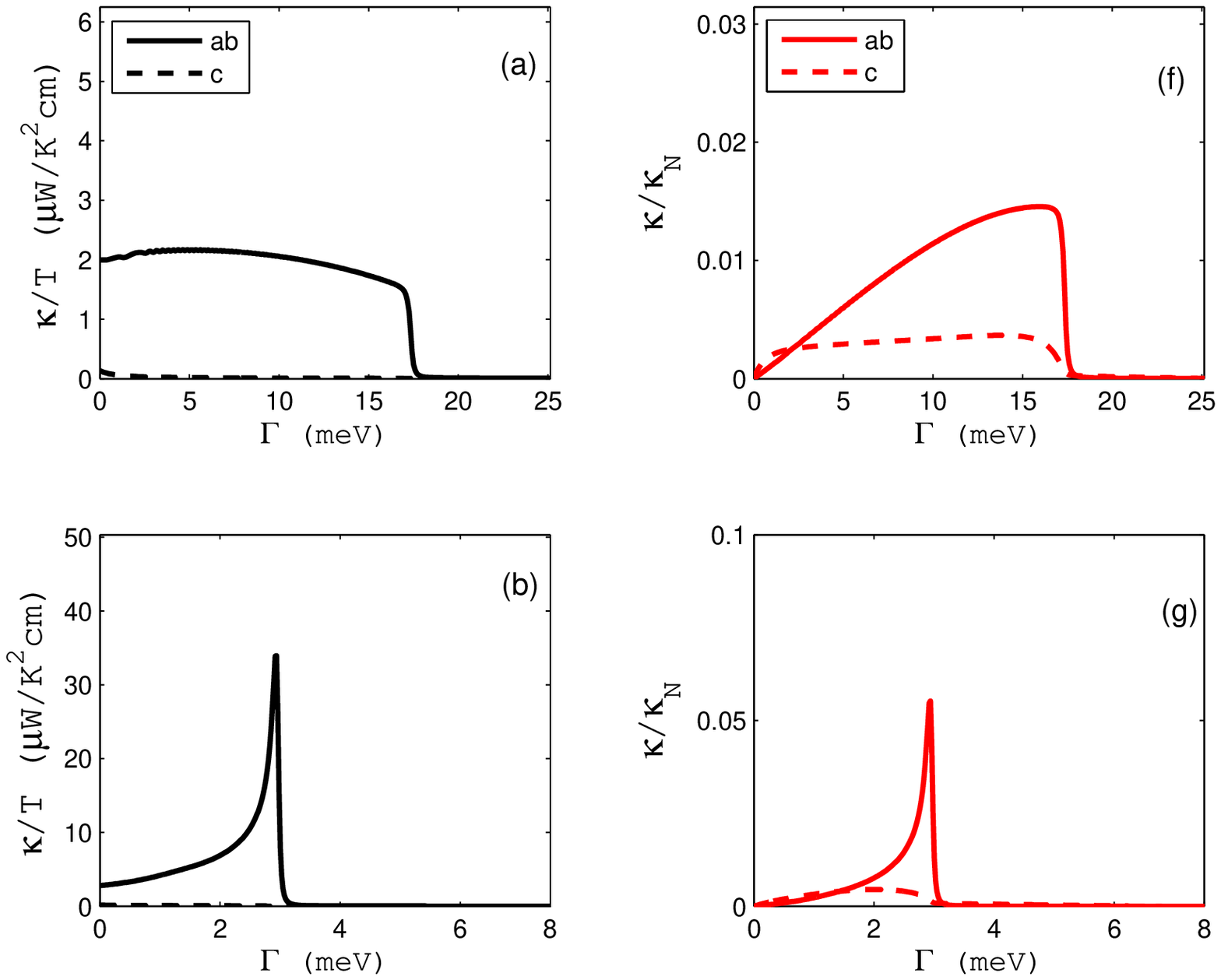}
\includegraphics[width=1.0\columnwidth]{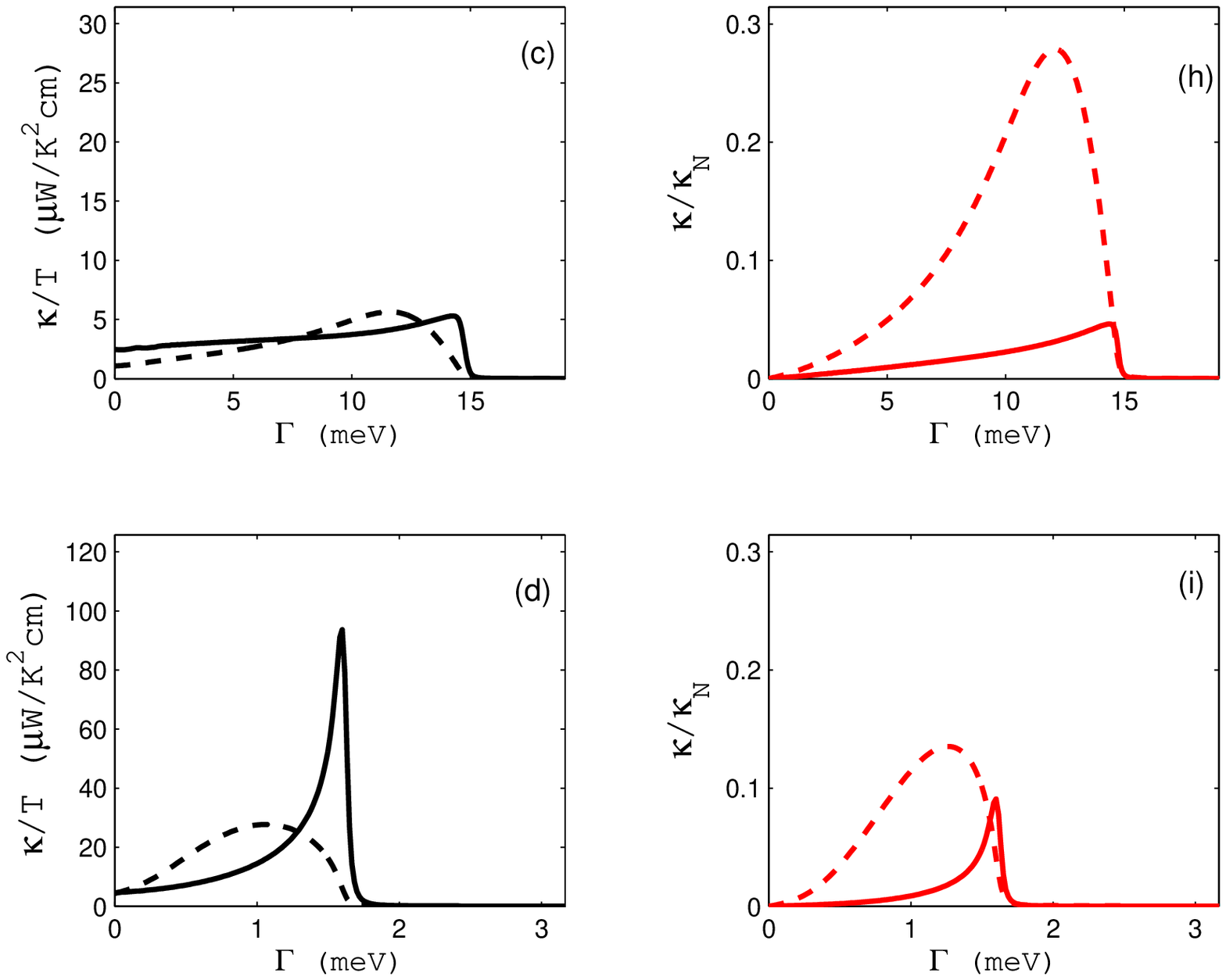}
\includegraphics[width=1.\columnwidth]{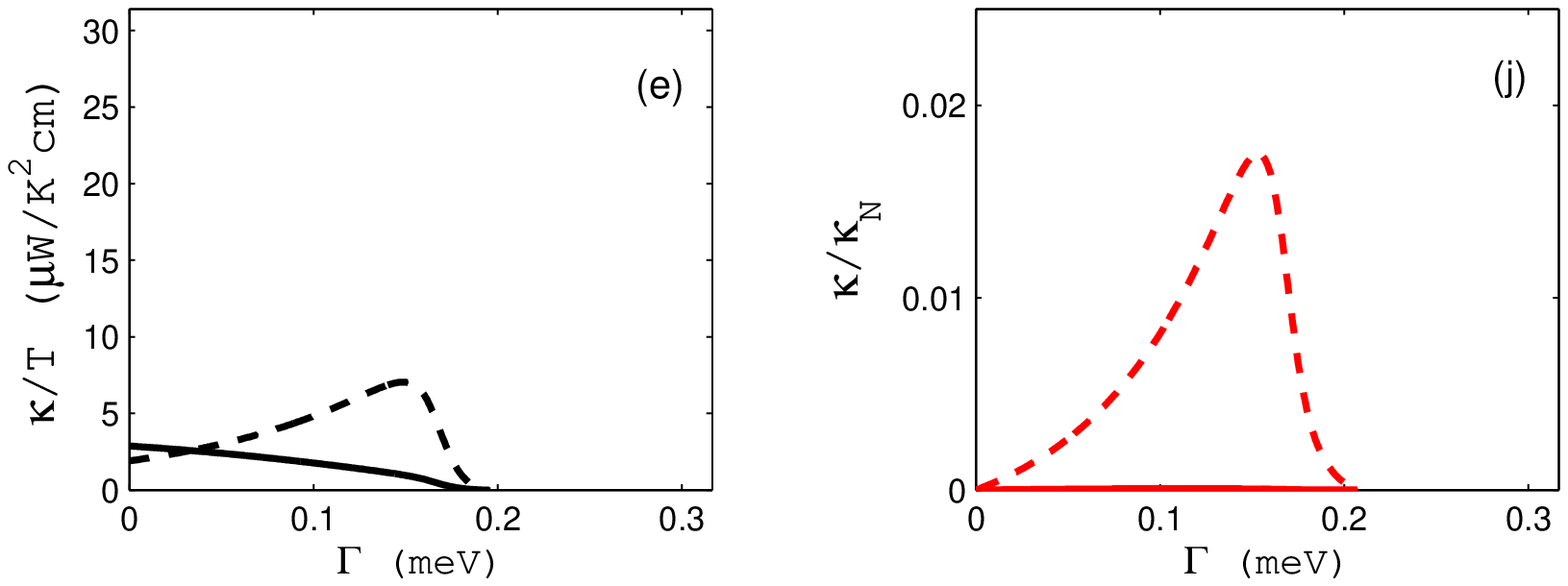}
\caption{Residual thermal conductivity $\lim_{T\rightarrow 0}\kappa(T)/T$ in units of $\mu$ W/K$^2$cm as a
function of  impurity scattering rate $\Gamma$ in meV.  Panels (a,c,e,g,i) correspond to absolute  thermal
conductivity for thermal current  in the $ab$-plane (solid) and along the  $c$-axis (dashed), respectively, for cases 1-5
  depicted in Fig. \ref{fig:FS}.   Panels  (b,d,f,h,j) correspond to normalized thermal conductivity $\kappa/\kappa_N$ for the same cases.  }
 \label{fig:kappa_abs}
\end{figure}

\begin{figure}
\hspace*{-5mm}
\includegraphics[width=0.95\columnwidth]{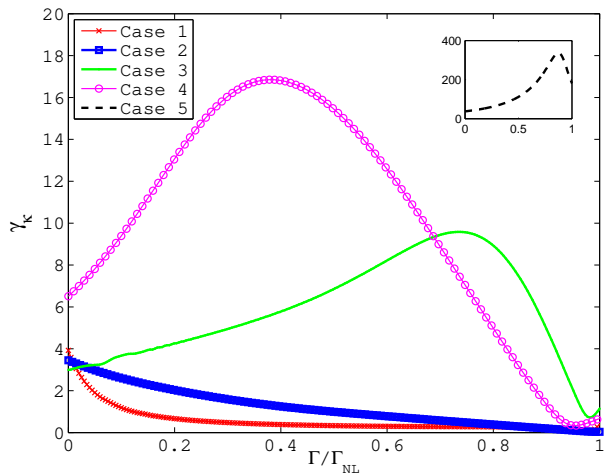}
\caption{Normalized thermal conductivity anisotropy ratio $\gamma_\kappa =\lim_{T\rightarrow 0}( \kappa_c/\kappa_{c,N})/( \kappa_{ab}/\kappa_{ab,N})$
as a function of the disorder scattering rate normalized to the critical value of disorder $\Gamma_\mathrm{NL}$ that lifts the nodes.}
 \label{fig:gamma_kappa}
\end{figure}

%
Depending on the Fermi surface, the corresponding $c$-axis transport can be smaller or larger than the $ab$-plane results discussed.
To illustrate  the qualitative anisotropy found here, we plot in Fig.~\ref{fig:kappa_abs} the $ab$-plane and $c$-axis thermal conductivities $\kappa/T$
for increasing  intraband impurity scattering. There are several interesting points in this Figure.  First, we note that for cases
1 and 2, where the Fermi surfaces are nearly cylindrical, the $ab$-plane conductivity is much larger than the $c$-axis conductivity.
Clearly again these cases are inconsistent with experiment. For the more flared Fermi surface, cases 3 and 4, it is possible
to have $\kappa_{ab}/T$ and $\kappa_{c}/T$ comparable, or even have $\kappa_c > \kappa_{ab}$.  Note this is more likely
for the horizontal line nodes of case 4, which have been chosen to lie at the maximum value of $v_{F,c}$ for this Fermi surface, whereas
the V-shaped nodes of case 3 sample the high $v_{F,c}$ values less often.  A large $c/a$ anisotropy is also observed in case 5 for reasons
similar to those of case 4.

As intraband scattering is increased, a constant term is added to the off-diagonal self-energy in Nambu space\cite{v_mishra_09} which eventually lifts
the nodes, as seen in the DOS (Fig.~\ref{fig:DOS}). As this occurs, two effects happen in parallel: a) $v_\Delta$ decreases; and  b)
the nodal surfaces (which are shown in Fig.~\ref{fig:FS} for the clean system only)
move towards the top and bottom of the Brillouin zone, respectively.  The latter effect
can  increase or decrease the thermal conductivity depending on whether or not the
node-lifting process moves the nodes towards regions of higher or lower $v_{F,c}$.  Thus, depending on
details, the $c$/$a$ anisotropy can be enhanced or suppressed by disorder.


In Ref.~\onlinecite{Reid2010}, the authors specifically discuss the {\it normalized} measure of thermal conductivity
anisotropy  $\gamma_\kappa\equiv \lim_{T\rightarrow 0}( \kappa_c/\kappa_{c,N})/( \kappa_{ab}/\kappa_{ab,N})$, and demonstrate that this
ratio has a minimum at optimal doping.  This result should  be treated with caution, however, both because a) the linear term in $\kappa_{ab}$
is consistent with zero for much of the doping range; and b) the normal state $\kappa_N$'s are determined
by measurement of the resistivity extrapolated to $T=0$ and use of the Wiedemann-Franz law.
However, if the elastic  transport scattering rate $1/\tau_{tr}$  is reasonably isotropic, as seems likely, $\gamma_\kappa$
will be determined simply by the ratio {\begin{eqnarray}(\langle v_{F,c}^2 \rangle_\mathrm{node}/\langle v_{F,c}^2 \rangle)/(\langle v_{F,ab}^2
\rangle_\mathrm{node}/\langle v_{F,ab}^2 \rangle),\label{eq:ratio_isotr}\end{eqnarray}
where $\langle v_{F,\alpha}^2\rangle_\mathrm{node}$ is the average of the Fermi velocity components $\alpha$ over
the nodal manifold.
} Here we calculate $\gamma_\kappa$ within a simple
model where normal state heat conduction is controlled by (isotropic) intraband impurity scattering only.  In Fig.~\ref{fig:gamma_kappa},
we show that only for cases 3-5 it is possible to achieve values of $\gamma_\kappa$ substantially greater than 1.
For case 5, the node lifting scattering rate $\Gamma_\mathrm{NL}$ is minuscule in absolute units, such that we rule out this case as well,
provided our assumption that intraband scattering dominates is correct.

An alternate possibility to create a small residual $\kappa/T$ would be the creation of an impurity band
in a gapped state due to interband scattering\cite{v_mishra_09}, {
but in this case there is no natural reason to have a  transport anisotropy other than that given
by $\kappa_c/\kappa_{ab}\sim\langle v_{F,c}^2\rangle/\langle v_{F,ab}^2\rangle$, which for all cases discussed here would give a
very strong dominance  by the $ab$-plane thermal currents on an absolute scale, and
would lead to $\gamma_\kappa \simeq 1$ in all samples, in contrast to experiment.}

\subsection{Magnetic field dependence of $\kappa$}

\begin{figure}
\hspace*{-.5cm}
\includegraphics[width=1.0\columnwidth]{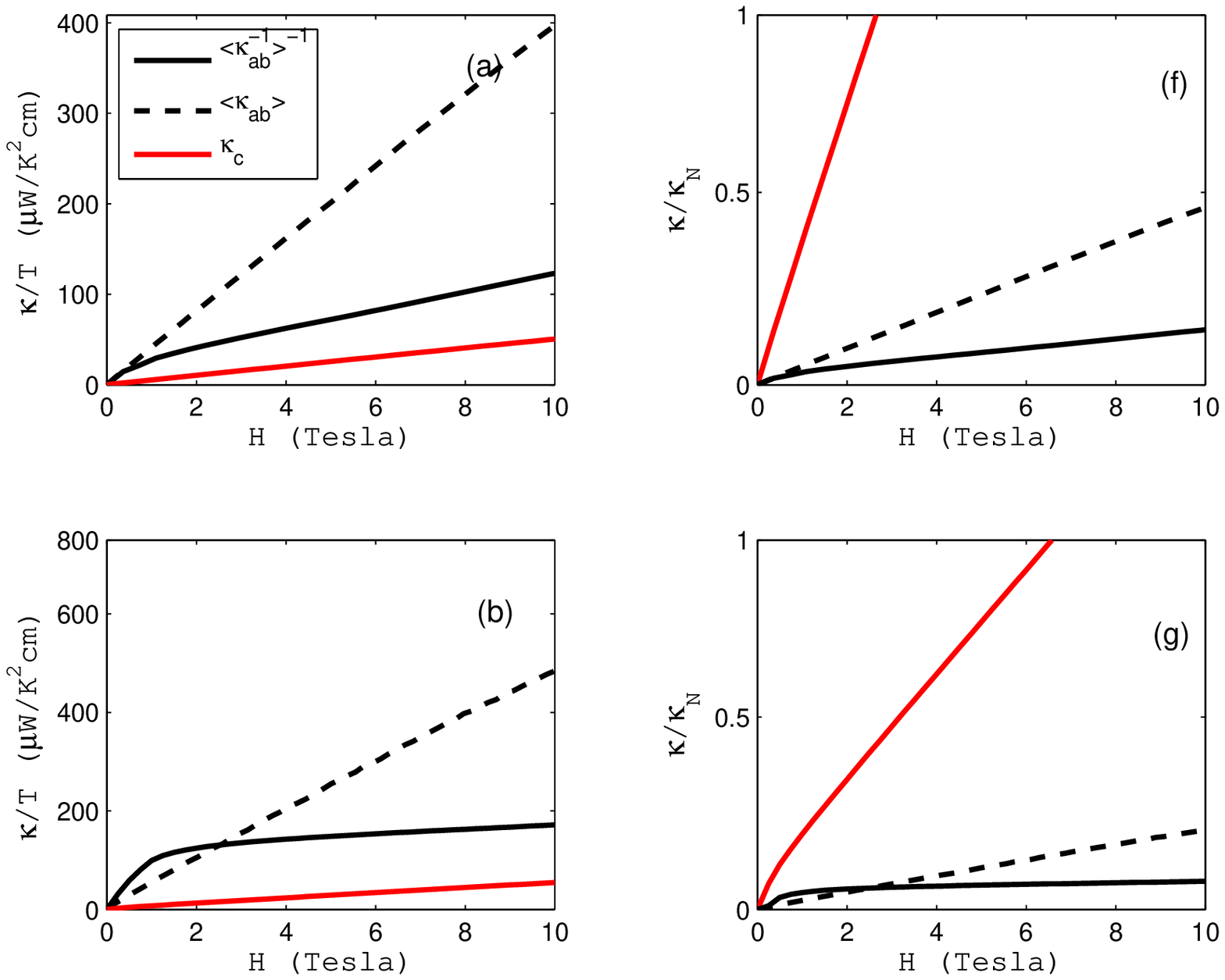}
\includegraphics[width=1.0\columnwidth]{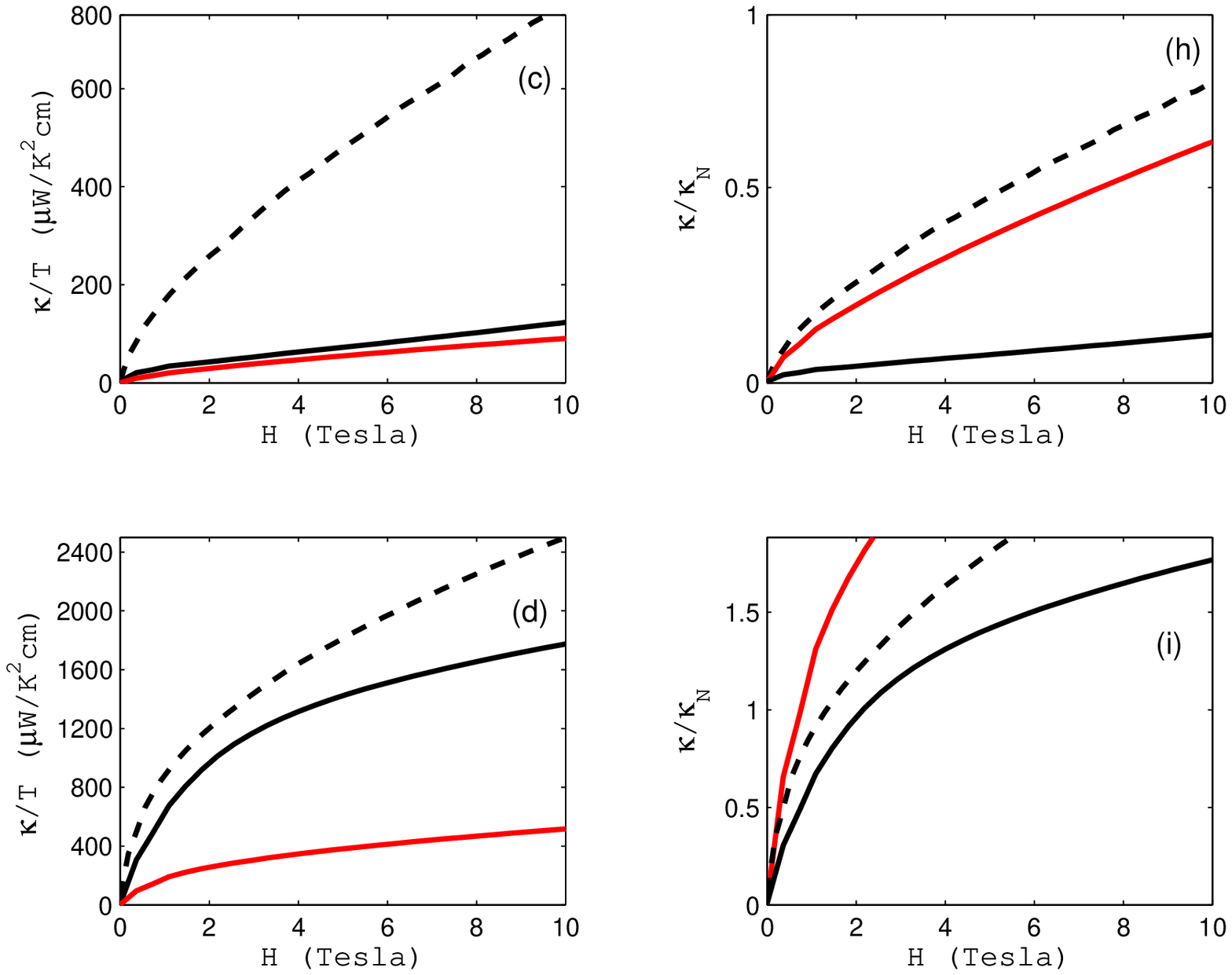}
\includegraphics[width=0.85\columnwidth]{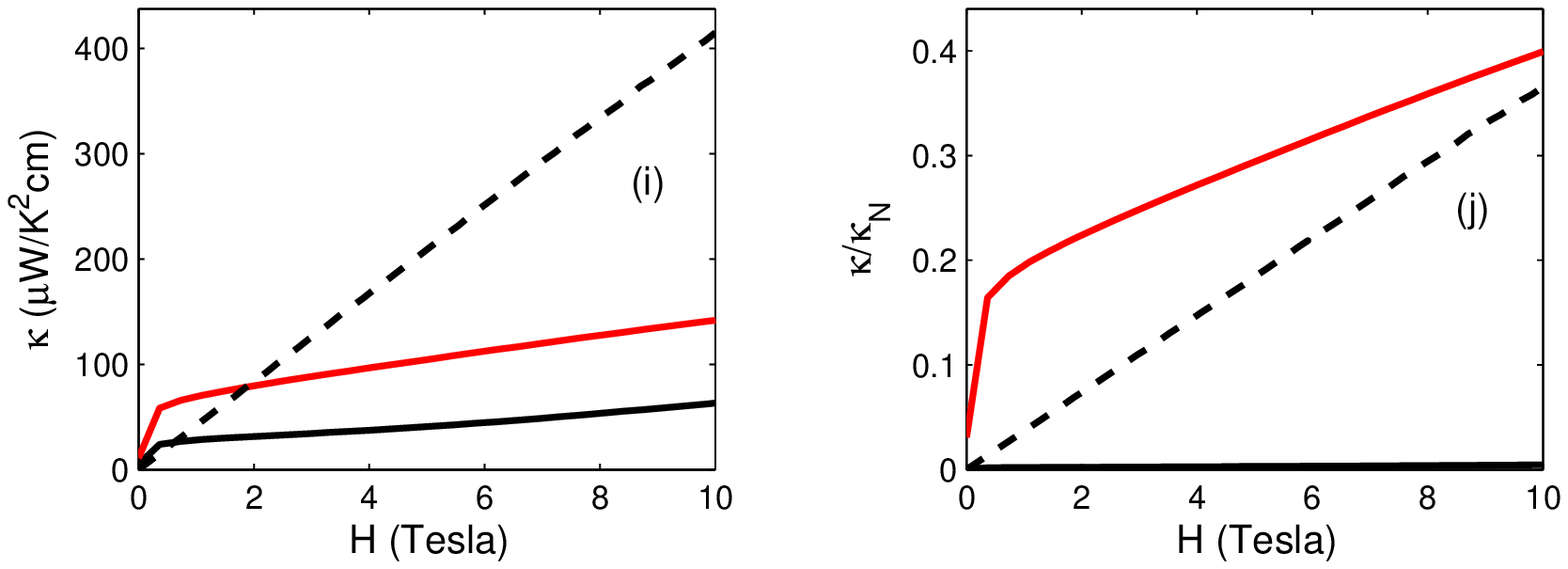}
\caption{ Residual thermal conductivity  $\kappa/T$ in units of $\mu$W/K$^2$cm vs.
magnetic field $H$ in Tesla.  Panels (a)-(e): cases 1-5 for Fermi surface $S_1$ with
intraband scattering rate $\Gamma$ = 1.3 meV (Case 1), $\Gamma$ = 0.6 meV (Case 2) and
 $\Gamma$ = 1.9 meV  (cases 3,4), and $\Gamma=0.06$ meV for case 5.}
\label{fig:kappaH}
\end{figure}

The final notable feature seen in experiment\cite{Tanatar2010,Reid2010} is an unexpectedly large enhancement
of the in-plane residual $\kappa/T$ term in the presence of a magnetic field along the $c$-axis. With moderate magnetic field of a
few Tesla, the residual in-plane term $\kappa_{ab}(H,T\rightarrow 0)/T$ becomes comparable to
the $c$-axis term $\kappa_{c}(H,T\rightarrow 0)/T$, and both exhibit a field dependence with downward curvature reminiscent of the
presence of field-induced residual quasiparticle excitations contributing to the thermal current\cite{c_kuebert_98}.
This effect was discussed already for the $ab$-plane thermal conductivity in Ref. \onlinecite{v_mishra_09}, where it was concluded
that the most likely origin of the strong field dependence in apparently fully gapped samples was contributions from deep minima in
the gap on the electron sheets, as found in early multiorbital spin fluctuation
theories\cite{wang2010,r_thomale_09,s_graser_08,k_kuroki_09,k_kuroki_08}.
Note that in the 3D spin fluctuation analysis of Graser \etal\cite{s_graser_10}, these deep minima on the electron sheets continue
to exist, and run the entire length of the Fermi cylinders, thus providing much more phase space than at zero field, as soon as the
magnetic field energy spans the very small minimum gap $\Delta_{min}$.  Thus the $ab$ thermal conductivity, {negligible} in zero field,
can jump fairly rapidly\cite{Mishra2009}.

To include the presence of magnetic field explicitly, in this work we use the standard semiclassical
(Doppler shift) method\cite{Volovik1993,c_kuebert_98,Vekhter1999,Vekhter2001}, where the quasiparticle energy
is modified by its Doppler shift in the local superfluid flow field.  While this approach is never exact\cite{m_franz_00},
it provides a good qualitative description of the magnetic field dependence of the quasiparticle properties of
nodal superconductors over a wide field range\cite{d_knapp_01}. The local Doppler shift of a quasiparticle with momentum
$\k$ (band indices are suppressed) is
\begin{equation}
\delta\omega (\k,\rr)={{{\bf v}_F(\k)\cdot {{\bf p}}_{s}(\rr)}}
 \label{eq:Doppler}
\end{equation}
where ${{\bf p}}_{s}(\rr)$ is the local supercurrent momentum, approximated for a single vortex as
\begin{equation}
{\bf p}_{s}(\rr)=\frac{\hbar}{2r}\hat{{\bf \theta}}
 \label{eq:Doppler2}
\end{equation}
Here $r$ is the distance from the vortex core and ${\theta}$ is the vortex winding angle. In the presence of the magnetic field,
we replace $\tilde{\omega}$ by $\tilde{\omega}-\delta\omega (\k,\rr)$ in Eq.~(\ref{eq:kappa1}).
In this approach, any physical quantity is calculated in the usual way with Doppler shifted energy, and the averaging
is performed over a unit cell of the vortex lattice, approximated as circular, as
\begin{equation}
\kappa(H)=\frac{1}{\pi R_H^2} \int_{0}^{R_H} dr \; r \int_{0}^{2 \pi} d\theta \kappa(\rr),
 \label{eq:doppler_kappa}
\end{equation}
where $\kappa(\rr,\theta)$ is the local (Doppler shifted) result from Eq.~(\ref{eq:kappa1}), $R_H = \sqrt{\Phi_0 / \pi H}$
is the magnetic length scale, with $\Phi_0$ the superconducting flux quantum. For lower fields $H\ll H_{c2}$, this
is a qualitatively good approximation. Somewhat more sophisticated approximations can be obtained within the
semiclassical framework\cite{Vekhter2001}, but the above will suffice for our qualitative purposes.
Here $H_{c2}\equiv {\Phi_{0}}/({2\pi \xi_{0}^{2} })$ is the upper critical field and $\xi_{0}= {v}_{F}/(2\pi T_c)$ is the coherence length.

Using this method, we now calculate the low-$T$ residual thermal conductivity in both directions as a function of {the magnetic} field.
Fig.~\ref{fig:kappaH} shows the field dependence of the residual linear term, which is similar to a superconductor with nodes,
for both directions of the heat current, with the magnetic field taken always to point along the $c$-axis as in experiment.
Note that in principle the situations with heat current parallel and perpendicular to the field are physically different
regardless of the orientation of the crystal axes: in one case the heat current flows down the axis of a vortex and may be
assumed to sample a single $\kappa(\rr)$, while in the other case it samples a modulated $\kappa(\rr)$ across the vortex lattice.
This averaging problem for the perpendicular heat flow has never been addressed in a completely convincing fashion.
We therefore present two extremes of the possible field dependence for currents in the  the $ab$-plane: first, a ``parallel average"
over the vortex flow field $\langle\kappa_{ab}\rangle \equiv \langle \kappa(\rr) \rangle$, where $\langle ...\rangle$ represents a
direct average over the vortex unit cell as in Eq.~(\ref{eq:doppler_kappa}); second, a ``series average", $\bar\kappa_{ab} \equiv
\langle \kappa(\rr)^{-1} \rangle^{-1}$, as discussed in Ref.~\onlinecite{c_kuebert_98}.  We note that the results for the Brandt-Pesch-Tewordt
average Green's function theory\cite{Mishra2009} appear to lie closer to the parallel average.

Fig.~\ref{fig:kappaH} shows several interesting effects.  First of all, while
the Fermi surfaces with less flaring (cases 1 and 2) displayed much larger values of  the absolute $ab$-plane thermal conductivity
in zero field {(Fig.~\ref{fig:kappa_abs}, panels (a) and (b))} than the more flared cases 3 and 4 {(Fig.~\ref{fig:kappa_abs}, panels (c) and (d))},
the $ab$-plane field dependence $\kappa_a(H)$ at any reasonable field is seen to dominate the absolute $c$-axis response
$\kappa_{c}(H)$ for all cases 1-4 {(Fig.~\ref{fig:kappaH}, panels (a-d))}. This is because we have assumed, as in Ref.~\onlinecite{Mishra2009},
that the $S_2$ sheet has a deep gap minimum, spanned by the Doppler field energy $E_H$ at a field $H$ of only 1.2 T.
Note this is an average scale over the entire sheet; the dispersion of the Fermi velocity with $k_z$ leads to a smearing of any
feature associated with this scale. It is these long deep gap minima on $S_2$ (putative electron sheets) which dominate the field
response in this case, with subdominant contributions from the $S_1$ (putative hole sheets) above  this field range.
Because $\kappa_{cN} \ll \kappa_{aN}$, it is still possible to have a normalized thermal conductivity larger in the {$c$-axis direction
(Fig.~\ref{fig:kappaH}, panels (f) and (g))}.

\begin{figure}
\includegraphics[width=0.9\columnwidth]{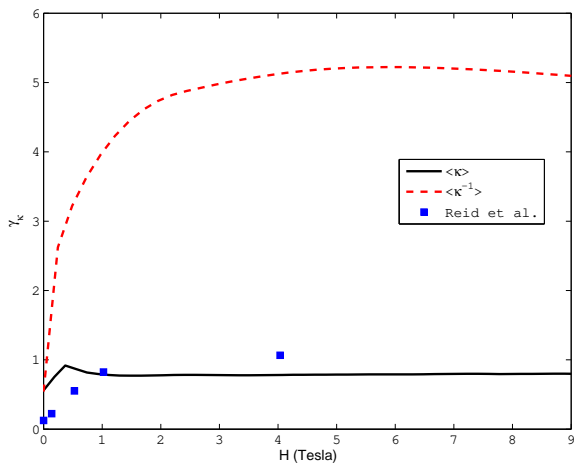}
\caption{Inverse normalized thermal conductivity anisotropy ratio $\gamma_\kappa^{-1} =\lim_{T\rightarrow 0}
(\kappa_a/\kappa_{a,N})/( \kappa_{c}/\kappa_{c,N})$  a function of magnetic field for Case 3.  Solid line: parallel
averaging $\langle \kappa_{ab} \rangle$; dashed line: series averaging $\langle (\kappa_{ab}^{-1}) \rangle^{-1}$
as described in text. Symbols: $\langle (\kappa_{ab}^{-1}) \rangle^{-1}$ data for highly overdoped $x=0.127$ sample of Ref. \onlinecite{Reid2010}, which  shows nonzero $\kappa_{ab}/T$ as $T\rightarrow 0$. }
 \label{fig:anisotropy_field}
\end{figure}

The Reid \etal~experiment Ref.~\onlinecite{Reid2010} shows a pronounced tendency for the normalized anisotropy
ratio $\gamma_\kappa$ to fall rapidly from large values as a function of
magnetic field, and then to saturate at values of order 1 when fields of a few Tesla are reached.
Within our calculational scheme, such a conclusion is quite natural since the zero-field anisotropy
(Fig.~\ref{fig:gamma_kappa}) $\gamma_\kappa$ is dominated by the weak nodes on the $S_1$ sheet(s), which
can give large $\gamma_\kappa$'s, particularly in cases 3 and 4, as discussed above. On the other hand, as the field is
switched on, the contributions of the deep  gap minima on the $S_2$ sheets come into play, and indeed dominate for sufficiently
large fields because of the much larger phase space  of the gap minima on $S_2$ compared to the phase space
taken by the gap  nodes on $S_1$. If a single 2D sheet dominates the field dependence, the normalized anisotropy
ratio will reduce to approximately 1 since the Fermi velocity anisotropy has been removed by the normalization,
{\it provided} the averaging over the vortex lattice is performed identically for in-plane and $c$-axis currents.
This behavior is seen in Figure~\ref{fig:anisotropy_field},
where the inverse of the anisotropy ratio, $\gamma_\kappa^{-1}$ is plotted to facilitate comparison with experiment.
If series-type averaging is more appropriate in the $ab$-plane, the high-field $\gamma_\kappa^{-1}$ can be considerably larger,
as also seen in Figure~\ref{fig:anisotropy_field}.

\subsection{Doping dependence}

\begin{figure}
\includegraphics[width=0.9\columnwidth]{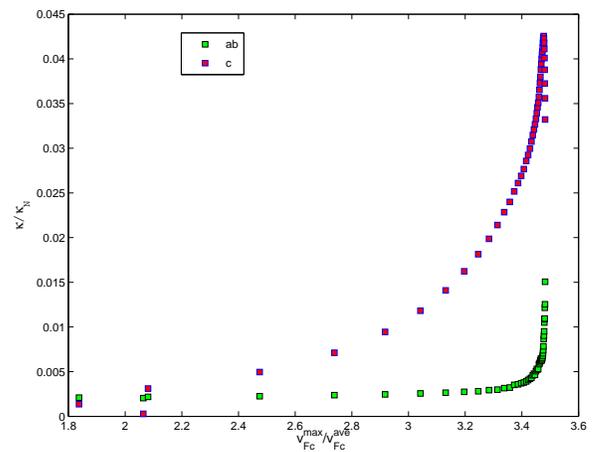}
\caption{Normalized thermal conductivity $\kappa/\kappa_N$ for heat current in $a$ and $c$ directions vs. degree of flaring of Fermi
surface $S_1$, parametrized by $v_F^{max}/v_F^{ave}$. }
 \label{fig:flaring}
\end{figure}

Within our current crude approach, doping is accomplished by rigidly shifting the chemical potential relative to the
parent compound.  However, the dependence of the Fermi surface thus obtained  is consistent neither with that observed in
ARPES experiments on 122 compounds nor with virtual crystal approximation within DFT calculations.
In particular, the flaring of the hole ($S_1$) Fermi surface upon electron doping does not increase rapidly with electron doping, as seen in
ARPES and DFT calculations\cite{Fink2009a,Brouet2009}.  In this section we therefore investigate, independently of any microscopic model of
the band structure, the effect on superconducting state transport of the flaring of the $S_1$ Fermi surface. To this end, we simply interpolate
the Fermi surfaces between cases (1,2) and (3,4), and calculate the thermal conductivity as outlined above. The superconducting order
parameter is held fixed in this process.
Within our current interpretation of the data on Co-doped Ba-122, increasing the amount of flaring is equivalent to doping the
system, although the correspondence between the two scales is nonlinear and not completely determined.  Nevertheless, it is
clear that an essential feature of the experiment, the strong increase of the normalized anisotropy ratio $\gamma_\kappa$ can be
understood by making this simple assumption.   Note that the normalization automatically divides out the average effect of the increasing
$v_{F,c}/v_{F,ab}$; what happens, therefore, is that the flaring increases particularly near the nodes and therefore preferentially enhances
low-energy $c$-axis quasiparticle transport.  Fig.~\ref{fig:flaring} illustrates the effect of the increased flaring, parameterized by the
ratio of the Fermi velocity components, for the order parameter with V-shaped nodes as in cases 1,3.  The effect is superficially quite
similar to the effect of doping on the anisotropy ratio observed by Reid \etal\cite{Reid2010}.

\subsection{Modulations of $\kappa$ with respect to magnetic field  angle}
\label{sec:polar}

Measurements of the specific heat and thermal conductivity as a function of the angle  between an
applied magnetic field and the crystal axes are in principle
powerful methods to determine the location of gap nodes~\cite{Vekhter1999,Matsuda2006}.
Since these measurements are probes of the bulk properties, and { the treatment of surfaces}
in the Fe-pnictide systems continue to pose problems, rotating field measurements were proposed early on as appropriate ways to map
nodal structures of the order parameter  in these systems~\cite{Graser2008A}. Measurements of the low temperature specific heat
as a function of the angle of an in-plane magnetic field have been reported very recently \cite{Zeng2010}.
While this method is sensitive to vertical line
nodes, it misses horizontal nodes altogether and can also miss nodal structures near the extrema
of the Brillouin zone for a flared Fermi surface.  Similar considerations apply to the oscillations of
the thermal conductivity with field angle, which historically have been more sensitive to weak nodal structures.
Here
we have calculated, for an arbitary direction of the magnetic field,
 the thermal conductivity for a fixed amplitude of the magnetic field for case 3.
\begin{figure}
 \includegraphics[width=1.0\columnwidth]{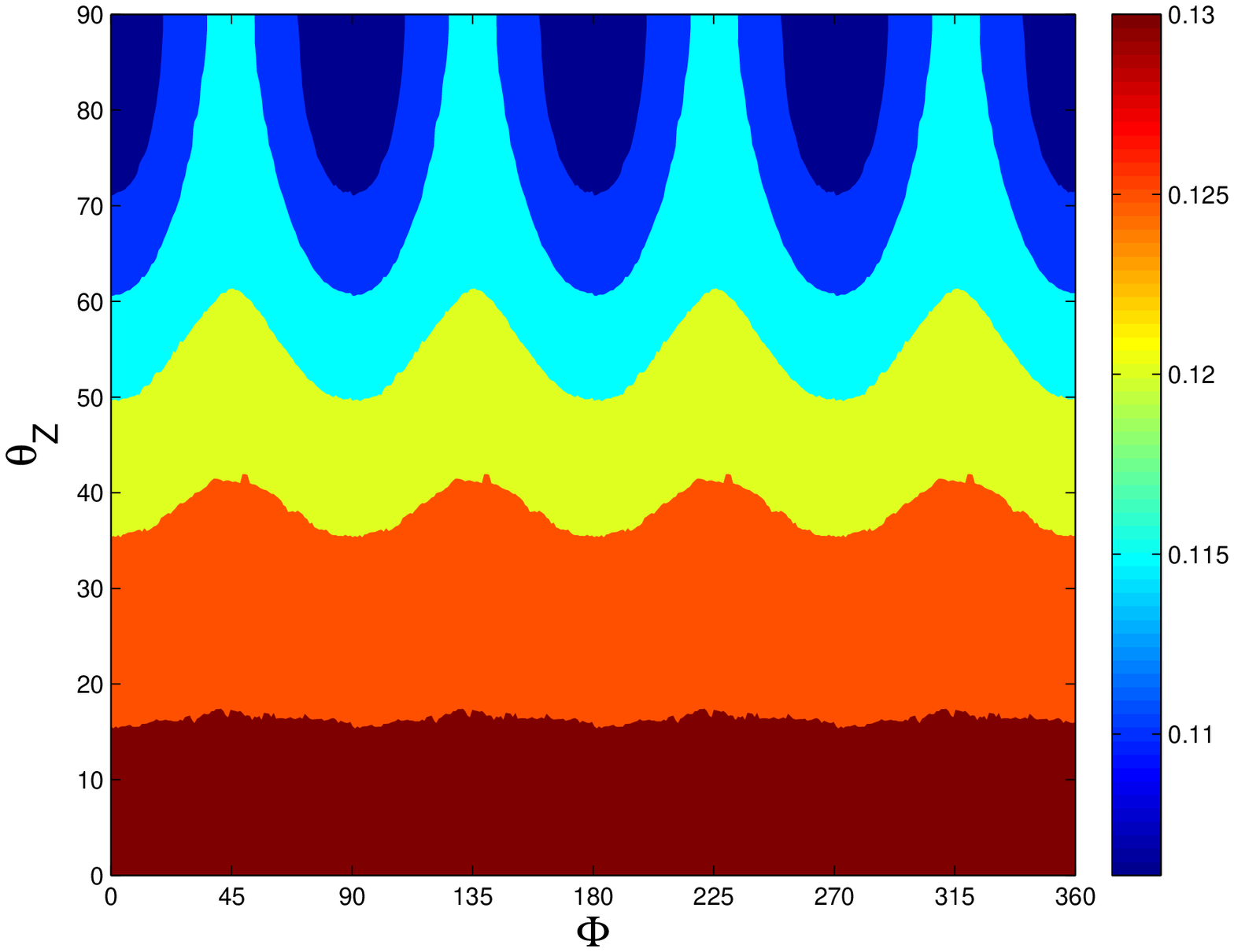}
\includegraphics[width=1.0\columnwidth]{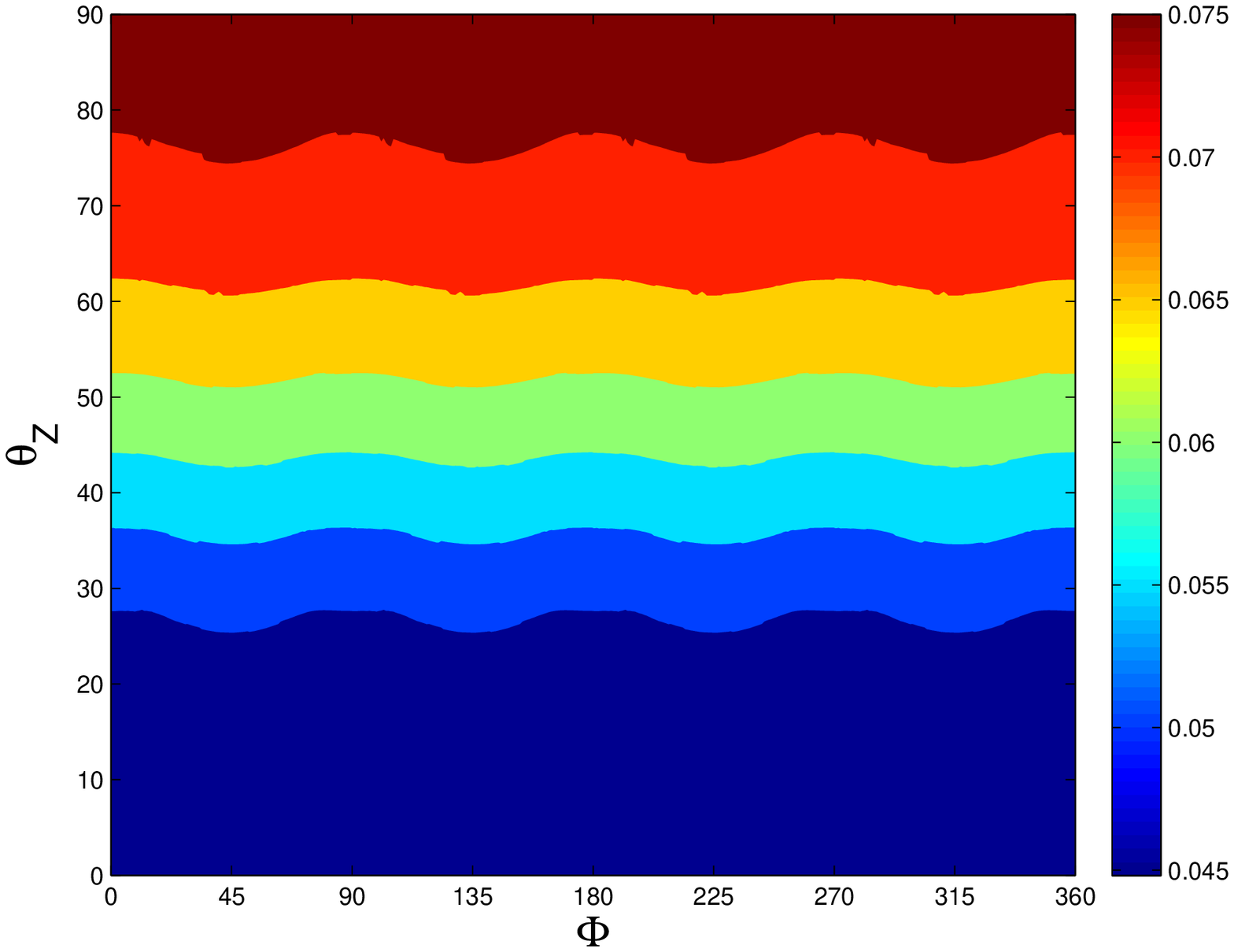}
\caption{Residual specific heat $C({\bf H})/C_N$ (top) and $c$-axis  thermal conductivity (bottom) $\kappa_c({\bf H},T\rightarrow 0)/\kappa_{cN}$ normalized to zero field normal state
values as a function of direction of applied magnetic
magnetic field described by polar angle $\theta_Z$ and azimuthal angle $\Phi$. The Fermi surface
and form of order parameter used for this calculation is same as case 3 with $\Delta_{h}=0.3meV$ and $\Delta_{e}=1.3meV$. Magnetic
field strength is $10.0T$. }
\label{fig:3D_field_dep}
\end{figure}

The results of these calculations are shown in Fig.~\ref{fig:3D_field_dep}. {The situation} for polar angles $\theta_Z$ near zero, i.e. for an alignment of
the field along the $c$-axis direction and the heat current applied with an in-plane angle $\Phi$,
clearly causes negligible thermal conductivity oscillation. As the polar angle is increased,
the expected characteristic fourfold oscillations emerge\cite{Vorontsov2006}.  Since we have evaluated the
conductivity at $T\rightarrow 0$, minima as a function of $\Phi$ for constant $\theta_{Z}$ reflect the existence of nodes at angles
$90^\circ -\theta_{Z}$. We have evaluated the $T$ dependence as well, and find that inversions of this pattern can occur at higher $T$
\cite{Vorontsov2006,Vorontsov2007,Boyd2009}, but have not focussed on this aspect here. Within our model, for most angles the observed
fourfold oscillations with minima at $0^\circ$ and $90^\circ$, correspond to the deep minima of the gap function on the $S_2$ sheet assumed
in our model.  Only when $\theta_{Z}$ approaches 90$^\circ$ are the nodal surfaces on $S_1$  probed by the field.  One should expect for
cases 1 and 3 (V-shaped nodes)  new double minima on both sides of $\Phi=$0$^\circ$, 90$^\circ$, etc. to appear due to the V-shaped
nodes {for polar angles} $\theta_Z \lesssim 90^\circ$. These are not visible in the Figure, however,
in part because the anisotropy of the Fermi surface masks small nodal features.
In our model it appears difficult to observe these small features, but it may nevertheless be possible to detect them
with high experimental resolution.   Since the cases with full horizontal
nodes are robust, one might think that a polar angle scan of the magnetic field would easily pick up minima due to these features,
but this effect, while present, is obscured in our calculations by the $c$-axis dispersion of the Fermi velocity.  Our study suggests
that a careful mapping of the weak $S_1$ nodal structures with this method, while clearly desirable, may unfortunately be difficult both due to
the small size of the related signals and of course the experimental implementation of the 3D field rotation.

\section{Conclusions}
\label{sec:conclusions}

In the presence of an  order parameter with 3D structure on an anisotropic FS, transport properties
may be qualitatively very different along the plane and the c axis, and  FS
topology may also play an important role. In particular, the Fermi velocity is a characteristic
weight factor for transport coefficients: a larger Fermi velocity
near the high density of quasiparticles around gap nodes will greatly enhance the superfluid density, microwave or
thermal conductivity.   Our calculations, within
the framework of anisotropic $s_\pm$ states found in recent spin
fluctuation calculations\cite{s_graser_10} for 122 systems, support the suggestion
of Reid \etal~that the anisotropy observed in the thermal conductivity at low $T$ in 122 systems is most likely due to the
occurrence of gap nodes on portions of the Fermi surface  with large $c$-axis component of the Fermi velocity. It is noteworthy
that from the point of view of microscopic theory, the nodal structures which dominate the $c$-axis transport in the {Co-doped Ba-122 materials} are
most likely to be found on the hole rather than electron Fermi surfaces, in contrast to  calculations appropriate for 1111 and other
more 2D materials  which found the opposite. {It is also possible that changes in microscopic 
 interaction parameters can lower the gap minima on the electron sheets so as to become nodes,
 as may occur in the P-doped 122 material. } Our calculations are phenomenological in nature and have only assumed that
the nodes occur on one sheet of either type.

Linear-$T$ terms in the thermal conductivity can arise whenever a finite density of residual quasiparticle states exists at
the Fermi level due to a small amount of disorder.  Some suggestions have been made that experiments indicating low-energy quasiparticle
states can be understood by considering an isotropic $s_\pm$ state without nodes but with interband scattering, such that an impurity
band is created at the Fermi level\cite{Kim2010a}. At least in the 122 systems, this possibility
now appears to be conclusively ruled out, for several reasons. First, such an impurity band would lead to a dominant $T^2$ dependence
of the penetration depth, not linear-$T$,  as reported by Ref. \onlinecite{c_martin_10} in 122 systems). Secondly, at low $T$ disorder
scattering is the dominant scattering mechanism and also determines the characteristic transport time. Calculations with realistic
impurities show weak variation of the relaxation rate around the Fermi surface; and the density of impurity band states is therefore
roughly isotropic; thus transport anisotropy will be determined in such a situation entirely by the ratio
$\langle v_{F,c}^2\rangle/\langle v_{F,ab}^2\rangle\ll 1$, rather {than} by the Fermi velocity values on special parts of the Fermi surface.
This result is now inconsistent with Ref. \onlinecite{Reid2010}.  For this reason we have neglected a treatment of interband impurity
scattering in this work, but realistic calculations will require a knowledge of both intra- and interband scattering amplitudes of
the dopant impurities.  

{In our work here we have also neglected magnetic impurities, in part because
the microscopic calculations of Kemper et. al. \cite{a_kemper_09}  reported a weak magnetic potential component of the Co dopant.  In addition, magnetic impurities are generically
strong pairbreakers in $A_{1g}$ states.  Similar to the discussion of interband nonmagnetic scatterers above, if they were strong enough to form impurity bands on the Fermi level and thus influence transport, they would create significant $T_c$ suppression, large values of the linear residual term in $\kappa_{ab}$,  and
a normalized anisotropy $\gamma_\kappa$ in zero field close to unity, none of which are observed in experiment.}

We have focussed here primarily on two properties, penetration depth and residual thermal conductivity in the $T\rightarrow 0$  limit.
For two concrete examples of Fermi surfaces ``inspired" by forms which occur in density functional theory calculations for certain 122 systems,
we have shown that one can simultaneously explain the existence of  large penetration depth  and thermal conductivity anisotropies
$\gamma_\lambda,\gamma_\kappa >1$, indicating a predominance of nodal quasiparticles with large $c$-axis components of the Fermi velocity.
Because of the complicated multiband nature of the system, as well as because the order parameters assumed have important low-energy
minimum gap scales, asymptotic low $T$ power laws will rarely be observed in the penetration depth $\Delta \lambda (T)$
if the $T$ dependence is fit to an experimentally relevant range.  We have therefore tried to give fits $\Delta \lambda (T)\sim T^\alpha$
to our theoretical results over such a finite range, and found  for two promising cases with ``flared" or closed Fermi surfaces in-plane exponents
$\alpha_{ab}\gtrsim 1.5-2$ which depend on details of order parameter structure and on topology of Fermi surface. On the other hand, the $c$-axis
penetration depth change is very close to linear $T$ dependence for these cases.   These results are quite close to those obtained in experiment
by Martin \etal\cite{c_martin_10}. Quantitative comparisons appear to us to be impossible at the present time, however, due to uncertainties
in the magnitudes of the zero $T$ penetration depths.

In the case of thermal conductivity, only the coefficient of the linear $T$ term is relevant since it can be unambiguously assigned
to electronic transport. We find again in the two promising cases that qualitative thermal conductivity anisotropy in agreement with experiment
is found. For this calculation,  semiquantitative agreement with experiment can be obtained once the disorder scattering rate and
overall Fermi velocity renormalization is fixed. One further remarkable aspect of the experiments  explained  by the 3D Fermi surface and
order parameter models we have considered is that the linear-T ab-plane thermal conductivity is substantially smaller than one might expect on the
basis of simple dimensional analysis, reflecting the small phase space from the short nodal segments near the top of the Brillouin zone.
The $c$-axis measurements of Reid \etal\cite{Reid2010}, in this picture, force one to the conclusion that $\kappa_{ab}$ is not zero
in these materials, as previously reported, but must be so small as to be unresolvable within current experiments.

Further conclusions can be drawn from study of the magnetic field dependence of the residual linear terms within the semiclassical approach.
The observed field dependence in the 122  is quite similar to nodal systems in clean limit, including the size of the term.  This suggests
that the small nodal structures probed by the $c$-axis transport are not giving rise to the field dependence; instead,
this contribution is dominated by the remaining Fermi surface sheets which have no nodes but rather deep gap minima which run the
length of the Fermi cylinder~\cite{v_mishra_09}. An immediate consequence is that in the cases which appear promising relative to experiment
in zero field, the $ab$-plane transport is immediately enhanced upon application of a small field, leading to comparable
nodal-type field variation of the thermal conductivity in both directions, as reported in Ref. \onlinecite{Reid2010}.

Finally, we have briefly discussed how the nodal structures which appear to us to be the most likely in this 3D system
could be detected.  We propose a full angle-dependent measurement of the directional thermal conductivity.  Weak 4-fold oscillations
of the specific heat  have been reported in other Fe-based superconductors\cite{Zeng2010}, and discussed theoretically recently.
\cite{Vorontsov2010} On the other hand, if the heat current is along $c$ and the field is
always in the plane, the unique features of the nodal structures near the top of the Brillouin zone face will be missed. Therefore
measurements at polar angles of the magnetic field between 0 and 90 degrees with respect to the plane may be useful.
Our study has found, however, that the {angular} dependence of this signal is probably dominated by the structures on
electron, not hole sheets; therefore these small hole pocket signatures may be difficult to observe.

Our calculations provide further evidence for the conjecture of an ``intrinsic sensitivity" of the gap structures in the
Fe-based superconductors to small perturbations in electronic structure~\cite{a_kemper_10}.  Recall that according to Graser \etal\cite{s_graser_10},
the nodes which appear near the top of the Brillouin zone on the hole pockets in these systems are probably related to the
suppression of $d_{xz}/d_{yz}$ orbital weight there, which in turn occurs because of special features in the 122 crystal structure
not shared by other Fe-pnictides.

Within our model, the increase of normalized thermal conductivity anisotropy observed by Reid \etal\cite{Reid2010} in Co-doped Ba-122
may be attributed to the increased flaring of the Fermi surface sheet as seen in ARPES experiments and DFT calculations.
This is a nontrivial effect due not to the increase of the average $c$-axis Fermi velocity but the increase of
this component at the position of the gap nodes.
There are clearly other effects present, however, including the possible occurrence of nodes on the electron sheets
for the most overdoped systems, as indicated by the non-zero $\kappa_{ab}$.  It is an important challenge to construct a microscopic
theory which can simultaneously describe the evolution of the Fermi surface and gap structure with doping.

\acknowledgments{
The authors would like to thank C. Martin, R. Prozorov, D.J. Scalapino, L. Taillefer, M. Tanatar, and  I. Vekhter  for helpful
comments and discussions. This work is {supported} by DOE DE-FG02-05ER46236 (PJH) {and the German Research Foundation through TRR80 (SG).}}


\end{document}